\documentclass[a4paper,11pt]{article}
\pdfoutput=1 

\usepackage{jheppub}
\usepackage{braket}
\usepackage[T1]{fontenc} 
\usepackage{amsmath, amsthm, mathtools}
\usepackage{hyperref}
\usepackage{filecontents}
\usepackage{jheppub} 
\usepackage[english]{babel}
\usepackage{mathrsfs}
\usepackage{slashed}
\usepackage{amssymb}
\usepackage{setspace}
\usepackage{tabu}
\usepackage{calligra}
\usepackage{wedn}
\usepackage{esint}
\usepackage{makecell}

\def\CA{{\mathcal A}}
\def\CF{{\mathcal F}}
\def\CL{{\mathcal L}}
\def\CM{{\mathcal M}}
\def\CN{{\mathcal N}}
\def\CO{{\mathcal O}}
\def\CP{{\mathcal P}}
\def\CR{{\mathcal R}}
\def\CS{{\mathcal S}}
\def\AdS{{\mathrm{AdS}}}
\def\CFT{{\mathrm{CFT}}}

\newcommand{\OGG}{ \overset{(2)}{\gamma} }
\newcommand{\OG}{ \overset{(0)}{\gamma} }

\title{\boldmath A note on holographic renormalization for non-conformal branes by recursion}

\author{Georgios Korpas}
\affiliation{Department of Computer Science, \\ Czech Technical University in Prague,\\Karlovo nam. 13, Czech Republic}

\abstract{
We review holographic renormalization for non-conformal branes using the Hamilton-Jacobi formalism. We provide the tools required for the computation of the holographic dictionary,  while we present some marginally new technical results regarding holographic renormalization for $Dp$-branes, $p<5$, for Einstein-Maxwell theory coupled to a dilaton. The computations are based on a recursive algorithm by Papadimitriou, that solves the radial Hamilton-Jacobi equation in the asymptotic boundary of the bulk. This paper is a short version of the Master thesis of the author written in 2014 at the Instituto de F\'isica Te\'orica, UAM, Madrid. 

}

\begin{document} 
\maketitle
\flushbottom

\section{Introduction}

The AdS/CFT correspondence \cite{Maldacena_1999}, or more generally the notion of gauge/gravity duality, is largely responsible for the modern understanding of the relation between the two pillars of modern physics, quantum field theory (QFT) and (quantum) gravity. This correspondence essentially provides a (non-rigorous) duality between QFT and gravity by conjecturing that a gravitational theory in a $d$-dimensional space (the ``bulk''), is equivalent to a (usually strongly coupled) QFT in a $d-1$-dimensional space which forms the boundary of the bulk. The gauge/gravity duality is known to hold on certain geometric backgrounds with the most nominal example being that of the AdS space (times a compact space) whose dual, in certain cases, is indeed a conformal field theory (CFT). Manifestations of this correspondence, often termed as the ``holographic correspondence'' or the ``holographic duality'', become evident when considering the path integrals or correlation functions on either side of the duality wherein one is interested in constructing a ``holographic dictionary'' \cite{Witten_1999,Gubser_1998}.  This dictionary, once formulated, can provide information for the strongly correlated observables on the QFT side. Specifically, the dictionary provides a two-way map $\CO(x) \leftrightarrow \Phi$, where $\CO(x)$ denotes local, single-trace, gauge invariant operators, often referred to as a ``boundary operators'' and $\Phi$ refers to fields in the bulk.
Let $W[J]$ denote the generating functional of (connected) correlation functions of the form $\braket{O_1(x_1)\ldots O_n(x_n)}$ and let $S[\Phi]$ denote the on-shell action of the bulk theory. The standard holographic dictionary makes the following identification
\begin{align}
    W[J] \sim S[\Phi]\Big|_{\Phi \sim J}.
\end{align}
Here, $J$ denotes a source on the QFT side. Specifically, this amounts to identifying $S[\Phi]$ with $W[J]$ evaluated on solutions of the equations of motion in the bulk subject to Dirichlet boundary conditions on the asymptotic AdS boundary. This is consistent since we can allow the transformation $\mathcal{L} \to \mathcal{L} + J(x)\CO(x)$ by interpreting it as a change of the boundary
condition of certain fields at the boundary of the AdS space. Given this identification, one can naively think that proceeding to compute any local, gauge-invariant QFT observables from the bulk theory duals is straight forward, at least in principle. Soon, one encounters a number of difficulties, which, to some extent, are to be expected, and we discuss them in what follows.

Quantum field theories are known to suffer from UV divergences that appear in, e.g., composite operators. This is true even in conformal field theories. For example, consider the two-point function $\left\langle\mathcal{O}\left(x_{1}\right) \mathcal{O}\left(x_{2}\right)\right\rangle \sim \frac{C}{\left|x_{1}-x_{2}\right|^{\Delta}}$ of a CFT, where $\Delta$ is the scaling dimension and $C$ a normalization constant. This is only a bare correlator and is ill defined at $x_{1}=x_{2}$ since it is proportional to $\Gamma(d / 2-\Delta)$ which is not defined for $\mathbb{Z}_{<0}$\footnote{Some results on negative integer arguments for the Gamma function already exist in the literature \cite{Fisher2012SomeRO}
However, in this paper we will not study such implications.} and as result, some form of regularization is required, forcing $\left\langle\mathcal{O}\left(x_{1}\right) \mathcal{O}\left(x_{2}\right)\right\rangle$ to a scale-dependent function. Classically, this is not a problem. However, in a quantum theory, the scale symmetry is broken. Of course, this does not apply to certain theories protected by supersymmetry. Such an example is the $d=4$, ${\rm SU}(N)$, $\mathcal{N}=4$ SYM theory, the CFT dual of the prototypical example of the AdS/CFT correspondence, which is UV finite, has finite correlations for its fundamental fields and a vanishing $\beta$-function. Despite these nice properties, composite operators of the form $\left\langle\mathcal{O}\left(x_{1}\right) \mathcal{O}\left(x_{2}\right) \ldots \mathcal{O}\left(x_{n}\right)\right\rangle$ are not well defined if their divergencies are not treated, a requirement to compute sensible and finite quantities. Thus, one needs to apply textbook renormalization techniques to such theories and compute renormalized correlation functions $\left\langle\mathcal{O}\left(x_{1}\right) \mathcal{O}\left(x_{2}\right) \ldots \mathcal{O}\left(x_{n}\right)\right\rangle_{\text {ren. }}$. Often it is more efficient to perform renormalization in the dual gravitational theory (should that exist), where the appearance of divergencies is justified due to the infinite bulk volume, that is, IR divergences of the AdS space. This program, renormalization in the bulk, is often coined as ``holographic renormalization'' and is a standard approach for computing interesting observable functions within AdS/CFT \cite{Henningson1998,Balasubramanian_1999,deBoer:1999tgo,deHaro:2000vlm,Bianchi_2001,Bianchi_2002,Martelli_2003,Skenderis_2002,Papadimitriou:2004ap,Papadimitriou2005, Kim2015, Papadimitriou_2010,Papadimitriou_2011,BenettiGenolini:2016tsn}.

The original work by Maldacena, Ref. \cite{Maldacena_1999} (see Ref. \cite{Aharony:1999ti} for an early review), seeded the idea of gauge/gravity duality by considering $\AdS_5/\CFT_4$ by considering coincident probe $D3$-branes within type II string theory on $\AdS_5 \times \mathbb{S}^5$. The boundary of $\AdS_5$ is locally asymptotically isomorphic to the usual Minkowski space in 3+1 dimensions. Holographic renormalization for non-conformal branes, as pioneered in \cite{Kanitscheider_2008}, takes into consideration bulk spaces that are non-AdS bulk and non-CFT boundary theories.

More concretely, once extending the original duality with bulk space $\AdS_5 \times \mathbb{S}^5$ as obtained by the backreaction of coincident $D3$-branes by substituting for arbitrary $Dp$-branes, the resulting world-volume QFT is again dual to a gravitational theory on the near-horizon
background generated by these $Dp$-branes \cite{Itzhaki1998}. Specifically, it is a $(p+1)$-dimensional
${\mathrm{U}}(N)$ SYM theory which, in contrast to the case of the $D3$-branes, is not a CFT since it has a scale-dependent dimensionful coupling constant, which is related to the bulk theory's running dilaton. Not all hope is lost since, as shown in \cite{Kanitscheider_2008}, in such backgrounds there exists a frame where a generalized conformal structure \cite{Jevicki_1998,Jevicki_1998b,Jevicki_1999,Kanitscheider_2008} exists and where the near-horizon geometry is conformal to $\AdS_{p+2}\times \mathbb{S}^{p-8}$. Exactly due to this generalized conformal structure, one is then able to formulate an asymptotic Fefferman-Graham expansion \cite{Fefferman95} of the solutions of the theory and, in turn, the construction of a renormalized action from which one can derive the holographically renormalized $n$-point functions of the dual boundary field theory. Furthermore, the generalized conformal symmetry, encoded in the radial direction, which is transverse to the boundary of $\AdS_{p+2}$, is interpreted as the RG flow parameter of the dual field theory.

Since its inception, holographic renormalization, in general, has been a fundamental aspect of AdS/CFT, much like renormalization in QFTs. For example, in \cite{Gutperle:2022pgw} and in the context of codimension 2 defects living in a $d=6$ SCFT, the gravity dual is a $d=7$, $\CN=4$ gauged supergravity and holographic renormalization is used, by including certain covariant terms, to get a finite renormalized action in the bulk. Holographic renormalization is also fundamental in a recent work \cite{Penin:2021sry} on strongly coupled mass-deformed CFTs with gravity duals in ${\rm dS}_3$ space. In Refs. \cite{Cabo-Bizet:2017xdr,Ntokos:2021duk}, holographic
renormalization is used in the study of thermodynamics of AdS black holes.   Holographic renormalization is also fundamental in understanding boundary conditions in topological $\AdS/\CFT$ \cite{BenettiGenolini:2020kxj}.

Motivated by the importance of holographic renormalization overall, we provide an example-based, (approximately) stand-alone analysis of the building blocks for performing holographic renormalization via the radial Hamilton-Jacobi method using a recurscive algorithm by Papadimitriou \cite{Papadimitriou_2010,Papadimitriou_2011}. Several excellent (and lengthy) reviews of holographic renormalization already exist in the literature \cite{deBoer:2000cz,Skenderis_2002, Papadimitriou2016, Arefeva2019}. In particular, in Ref. \cite{Papadimitriou2016}, a detailed emphasis is given precisely in the Hamilton-Jacobi formulation of the problem as presented here. This dates back to the original idea of Ref. \cite{deBoer:1999tgo} to use the Hamilton-Jacobi method (in the bulk) in order to identify and separate from the gravitational bulk on-shell action terms that can be written as local functionals of the asymptotic boundary geometry. This idea was systematically developed in \cite{Martelli_2003,Papadimitriou:2004ap} with Papadimitriou prescribing a recursive algorithm \cite{Papadimitriou_2010} (see \cite{Papadimitriou2018} for an online lecture) that we follow closely here. This method has several advantages over the first attempts to perform holographic renormalization (for example, there is no bulk-covariance breaking). Additionally, the radial Hamilton-Jacobi method as proposed in Refs. \cite{Papadimitriou:2004ap,Papadimitriou2005,Papadimitriou_2010,Papadimitriou_2011} proved to be important in the sense that, at least in principle, it makes possible to algorithmically construct the reduced
phase space for any bulk model at the boundary. 
Renormalization with the radial Hamilton-Jacobi method has found a variety of applications (some of which were referenced earlier) including the holographic understanding of supercurrent anomalies \cite{Papadimitriou:2017kzw} in $d=4$, $\mathcal{N}=2$ SCFTs. We discuss further applications in the conclusions. The take home message is that this approach for holographic renormalization can be performed by reformulating the gravitational bulk dynamics in terms of a symplectic space of boundary information, which, in turn, enables us to identify them with the space of the renormalized observables of the dual field theory, achieving at
the same time to make our bulk problem well defined at infinity \cite{Papadimitriou_2010}, where otherwise the on-shell action is pathological.

\section{Holographic renormalization for the Einstein-Maxwell-dilaton theory}

We start our analysis by introducing the action for the system of interest system: $d+1$-dimensional Einstein-Maxwell-dilaton theory defined on a Riemannian manifold $\CM$ equipped with a metric tensor $g_{\mu \nu}$. The action reads as follows
\begin{align}\label{eq:action}
S = \frac{1}{2\kappa^2} \int_{\CM}  \sqrt{-g} \left(R - \frac{1}{2}(\partial \varphi)^2 - \frac{1}{4}e^{\alpha \varphi} F^2  - V(\varphi)  \right) + \frac{1}{2\kappa^2} \int_{\partial \CM}  \sqrt{-\gamma}2K. 
\end{align}
Here, $\kappa = 8\pi G_{d+1}$ is a constant, $G_{d+1}$ is the $d+1$-dimensional Newton constant, $V(\varphi)$ is the potential function, $\gamma$ is the induced metric on the boundary, $K$ is the trace of the extrinsic curvature in the boundary, and the boundary integral is the Gibbons-Hawking term \cite{Gibbons1977}. This action has the
pathology of a variational problem that is not well defined \cite{Papadimitriou_2010}. Our approach is to treat this problem in analogy to \cite{Papadimitriou_2010, Papadimitriou_2011}. Importantly, we will find the boundary terms required to be included such that the on-shell action is finite, a term which also makes the variational problem well defined.

\subsection{Hamilton-Jacobi equations}

Under the ADM decomposition \cite{PhysRev.116.1322} the bulk metric takes the form
\begin{align}
g=
 \begin{pmatrix}
   {N^2 + N^iN_i}        & &  {N_i} & \\
   {N_j}         & &   \gamma_{ij}  \
 \end{pmatrix},
\end{align}
where $\gamma$ is the spatially reduced by one dimension induced metric, $N$ is the lapse function and $N_i$ the normal, to the foliated slice, shift vector, which essentially are Lagrange multipliers for the newly formed constraind system.

Returning to the action \eqref{eq:action}, we can decompose it into three sector-related summands to make the computations easier to tackle. We will work with each piece separately and then join them to obtain the ADM decomposed action. The action at hand can be decomposed as 
\begin{equation}
S = S_G + S_{\varphi} + S_{A},
\end{equation}
where
\begin{align}
S_G &= \frac{1}{2\kappa^2} \int_{\CM}  \sqrt{-g}R + \frac{1}{2\kappa} \int_{\partial M} \sqrt{-\gamma}2K,\\[0.5em]
S_{\varphi} &= \frac{1}{2\kappa^2} \int_{\CM}  \sqrt{-g} \left( -\frac{1}{2} (\partial\varphi)^2  - V(\varphi) \right),\\[0.5em]
S_A &= \frac{1}{2\kappa^2} \int_{\CM}  \sqrt{-g}\left( - \frac{1}{4}e^{\alpha \varphi} F^2 \right).
\end{align}

Naively, one might think that these ``sub-actions'' are decoupled while they are really not. The gauge action $S_A$ is coupled to the dilaton term, and thus it cannot be simply considered to be decoupled from the scalar field action $S_{\varphi}$. Nevertheless, this subtlety will have no effect in the subsequent analysis and especially in the recursive algorithm to be presented later on. 

With the above in mind, we are able to write the Hamiltonian density with support on a constant-radius slice $\Sigma_r \subset \CM$ as follows
\begin{align}
H = \int_{\Sigma_r}  \left( N\mathcal{H} + N_i\mathcal{H}^i + \mathcal{F}a \right).
\end{align}
Here, $\mathcal{F}$ is a yet to be determined function while $a$ is the scalar-valued radial component of the gauge field.

The gravitational part $S_G$ has to be rewritten in terms of the Ricci scalar $R[\gamma]$ of the induced metric $\gamma_{ij}$. The precise relation between the two Ricci scalars is given by
 \begin{align}
R[g] = R[\gamma] + K^2 - K_{ij}K^{ij} + \nabla_{\mu}(-2Kn^{\mu} + n^{\rho}\nabla_{\rho}n^{\mu}), 
\end{align}
where $n^{\mu}$ is the unit normal vector to $\Sigma_r$ and $\nabla_\mu$ is the covariant derivative with respect to the bulk metric $g_{\mu \nu}$. The extrinsic curvature $K_{ij}$ of $\Sigma_r$ is given by
\begin{align}\label{eq:extrinsic}
K_{ij} = \frac{1}{2}( \dot{\gamma}_{ij} - D_iN_j - D_jN_i)
\end{align} 
where $D_i$ is the induced covariant derivative with respect to $\gamma_{ij}$, while the dotted quantities denote radial derivatives, that is, $\dot{f} \coloneqq \partial f/\partial r$. The action of the gravity sector then reads
\begin{equation}
S_G = \int dr \CL_G,
\end{equation}
with the Lagrangian $\CL_G$ iven as 
\begin{equation}
 \CL_G = - \int_{\Sigma_r} \frac{ \sqrt{\gamma} N }{ 2\kappa^2 } \Big(R[\gamma] + K^2 + K_{ij}K^{ij} \Big).
\end{equation}
A total derivative term $\nabla_{\mu}( -2Kn^{\mu} + \nu^{\rho}\nabla_{\rho}n^{\mu} )$ that should normally appear above is canceled by the Hawking-Gibbons term of Eq. \eqref{eq:action}. 

Next, we wish to compute the conjugate momenta of this Lagrangian with respect to the induced metric, that is,
\begin{align}
\pi_{\gamma}^{ij} = \frac{\delta \mathcal{L}_G}{\delta \dot{\gamma}_{ij}}.
\end{align}
The result is as follows.
\begin{equation}\label{eq:conj_mom_ij}
\pi^{ij} = - \frac{1}{2\kappa^2}\sqrt{\gamma}(K\gamma^{ij} - K^{ij}), \qquad \pi_{ij} = - \frac{1}{2\kappa^2}\sqrt{\gamma}(K\gamma_{ij} - K_{ij}).
\end{equation}
In the equation above and onwards, for ease of notation, we skip the $\gamma$ (or the corresponding bulk field) index for the tensorial momenta quantities. However, we keep the corresponding index $\varphi,A,\gamma$ when confusion might appear for contracted scalar momenta quantities.

By rearranging Eq. \eqref{eq:extrinsic} we obtain the following \emph{flow equation in the radial direction}
\begin{equation} 
\dot{\gamma}_{ij} = 2NK_{ij} + D_iN_j + D_jN_i,
\end{equation}
see \cite[Eq. (2.1)]{Papadimitriou_2011}. By acting from the left of both sides of Eq. \eqref{eq:conj_mom_ij} with $\gamma^{ij}$, that is, $\pi_{ij} \to \gamma^{ij}\pi_{ij}$, we get
\begin{equation}
\begin{aligned}
\gamma^{ij}\pi_{ij} &= -\frac{1}{2\kappa^2}\sqrt{-\gamma}(dK - K) \\
			    &=-\frac{1}{2\kappa^2}\sqrt{-\gamma}(d - 1)K. \
\end{aligned} 
\end{equation}
Next, solving for $K$ gives
\begin{align}\label{eq:K}
K = -\frac{2\kappa^2}{\sqrt{-\gamma}}\frac{1}{d-1}\pi.
\end{align}
Finally, we are able to solve for $K_{ij}$ using Eq. \eqref{eq:K} above. We have
\begin{equation}
\begin{aligned}
    \pi_{ij} &= -\frac{1}{2\kappa^2} \sqrt{-\gamma} \left( \frac{-2\kappa^2}{\sqrt{-\gamma}} \frac{1}{d-1}\pi \gamma_{ij} - K_{ij} \right) \\
            &= \frac{1}{d-1} \pi \gamma_{ij} + \frac{1}{2\kappa^2}\sqrt{-\gamma}K_{ij}, \
\end{aligned}
\end{equation}
and as a result we can express the extrinsic curvature in terms of the conjugate momenta of the gravitational section of our action as
\begin{equation}
K_{ij} = -\frac{2\kappa^2}{\sqrt{-\gamma}} \left( \pi_{ij} - \frac{\pi \gamma_{ij}}{d-1} \right).
\end{equation}

In order to use the canonical transformations as in Ref. \cite{Papadimitriou_2010,Papadimitriou_2011} we need to construct the corresponding Hamiltonian by taking the Legendre transform. Specifically,
we can take the product $ \pi^{ij}\dot{\gamma}_{ij} $ by using the equations above to find
\begin{equation}
    \begin{aligned} 
H_G &= \int_{\Sigma_r}   \pi^{ij}\dot{\gamma}_{ij} - \CL_G \\
   &= \int_{\Sigma_r}   ( N\mathcal{H}_G + N_i\mathcal{H}_G^i),
\end{aligned}
\end{equation}
obtaining two ``Hamiltonian contraints'' as follows
\begin{align} 
\mathcal{H}_G = -\frac{2\kappa^2}{\sqrt{-\gamma}} \left( \pi_{ij}\pi^{ij} - \frac{1}{d-1}\pi^2 \right) - \frac{\sqrt{-\gamma}}{2\kappa^2}R[\gamma]  
\end{align}
and
\begin{align}  
\mathcal{H}_G^i = (-2D_j)\pi^{ij}.  
\end{align}

A similar computation can be performed for the dilaton term of the action \eqref{eq:action}. Naturally, being a scalar field, the dilaton computations are quite straightforward. We perform the ADM decomposition for the action $S_\varphi = -\int dr \, \CL_\varphi$ using the inverse metric we have calculated and expand the terms to obtain
\begin{align}
2\kappa^2 \CL_\varphi =  \int_{\Sigma_r} \frac{1}{2}\sqrt{-\gamma}N\left[ \left( \dot{\varphi} + N^i\partial_i\varphi \right)^2 - \gamma^{ij}\partial_i\varphi\partial_j\varphi \right] + \int_{\Sigma_r}  V(\varphi).
\end{align}

The corresponding conjugate momenta are given by
\begin{equation}
    \begin{aligned}
\pi_{\varphi} &= \frac{\delta \CL_\varphi}{\delta \dot{\varphi}} \\
             &= -\frac{\sqrt{-\gamma}}{2\kappa^2N}(\dot{\varphi} - N^i\partial_i\varphi).
\end{aligned}
\end{equation}
The dilaton flow equation reads
\begin{align}
\dot{\varphi} = -\frac{2\kappa^2N\, \pi_{\varphi}}{\sqrt{-\gamma}}  + N^i\partial_i\varphi,
\end{align}
and the dilaton Hamiltonian constraints are
\begin{align}
\mathcal{H}_{\varphi} = \frac{-2\kappa^2}{\sqrt{-\gamma}} \frac{\pi^2}{2} + \frac{\sqrt{-\gamma}}{2\kappa^2} \left( \frac{1}{2} \gamma^{ij} \partial_i \varphi \partial_j \varphi + V(\varphi) \right),
\end{align}
and
\begin{equation}
\mathcal{H}_{\varphi}^i = \pi_{\varphi} \partial^i\varphi.
\end{equation}

Finally, we are left to treat the gauge field action $S_A$. Proceeding as before, utilizing the ADM decomposition, the field-strength term of the action can be written as
\begin{align}
F_{\mu \nu}F^{\mu \nu} = \frac{2\gamma^{ij}}{N^2} (F_{ri} + N^kF_{ik})(F_{rj} + N^kF_{jk}) + \gamma^{ij}\gamma^{lk}F_{il}F_{jk}.
\end{align}
The Lagrangian is given by
\begin{align}
\CL_A =  \frac{\sqrt{-\gamma}}{\kappa^2N}\Big( \gamma^{ij} (F_{ri} + N^kF_{ik})(F_{rj} + N^kF_{jk}) + \gamma^{ij}\gamma^{lk}F_{il}F_{jk} \Big)e^{\alpha \varphi},
\end{align}
and the conjugate momenta are given by
\begin{equation}
   \begin{aligned}  
\pi^i &=  \frac{\delta \CL_A}{ \delta F_{ri}} \\
	   &=  \frac{2\sqrt{-\gamma}}{\kappa^2N}( F_r^i + N^kF_k^i )e^{\alpha \varphi},
\end{aligned} 
\end{equation}
The gauge field flow equation obtained reads
\begin{equation}
\dot{A}^i = e^{-\alpha \varphi} \frac{\kappa^2N}{2\sqrt{-\gamma}} \pi^i + \partial^i a - N^kF_k^i,
\end{equation}
where we recall that $a$ corresponds to non-dynamical gauge field in the radial direction, a Lagrange multiplier as well.
The Legendre transform needed for the Hamiltonian requires to compute the product of the conjugate momenta with the velocities 
\begin{align}
\pi^i\dot{A_i} = -e^{-\alpha \varphi}\frac{\kappa^2N}{2\sqrt{-\gamma}}\pi_A^2 + \pi^i\partial_i a.
\end{align}
By subtracting the Lagrangian from the above equation, we obtain the Hamiltonian for the gauge sector
\begin{equation}
\begin{aligned} \nonumber
H_A &= \int_{\Sigma_r} (  \pi^i \dot{A}_i - \CL_A ) \\
       &= \int_{\Sigma_r}   (N\mathcal{H}_A - \mathcal{F}a),
\end{aligned}
\end{equation}
with the Hamiltonian constraint being
\begin{equation}
\mathcal{H}_A = \frac{-\kappa^2 \pi_A^2}{\sqrt{-\gamma}e^{\alpha \varphi}} + \frac{\sqrt{-\gamma}e^{\alpha \varphi}}{8\kappa^2}F_{ij}F^{ij}
\end{equation}
and the sought after function $\mathcal{F}$ providing a gauge constraint
\begin{equation} \label{CKV}
\mathcal{F} = -D_i\pi^i,
\end{equation}
concluding the derivation of the constraints required for performing holographic renormalization as a canonical transformation. To summarize and collect the important formulae,the Hamiltonian, momentum and gauge constraints for our system, respectively, are given by
\begin{equation}\label{constraint_hamiltonian}
    \begin{aligned}
    \mathcal{H} &= -\frac{2\kappa^2}{\sqrt{-\gamma}} \left( \frac{\pi_{\varphi}^2}{2}  + \frac{\pi_A^2}{2e^{\alpha \varphi}} + \pi_{ }^{ij}\pi_{ij}^{ } - \frac{1}{d-1}\pi_{\gamma}^2  \right) \\
    &\quad   +  \frac{\sqrt{-\gamma}}{2\kappa^2} \left( -R[\gamma] + \frac{1}{2}  \gamma^{ij} \partial_i \varphi \partial_j \varphi + V(\varphi) + \frac{e^{\alpha \varphi}}{4}F_{ij}F^{ij} \right), \\
\end{aligned}
\end{equation}

\begin{equation}\label{constraint_momentum}
    \begin{aligned}
    \mathcal{H}_i    =   (-2D^{j})\pi_{ij}^{} + \pi_{\varphi}\partial_i \varphi + \pi^j F_{ij} ,
    \end{aligned}
\end{equation}

\begin{align}\label{constraint_gauge}
    \mathcal{F}     = -D_i \pi_A^i. 
\end{align}

A remark is due. In a general covariant theory the Hamilton-Jacobi equation is equivalent to the vanishing our the constrains we have just derived. Therefore, we can use Eqs. \eqref{constraint_hamiltonian},\eqref{constraint_momentum} and \eqref{constraint_gauge} and solve them for different potentials $V(\varphi)$.

Finally, before moving to the next section, let us make the observation that the symplectic form $\Omega$ given by
\begin{equation}
\Omega = \int_{\Sigma_r}   \Big( \delta \pi^{ij} \wedge \delta \gamma_{ij} +  \delta \pi^i \wedge  \delta A_i +  \delta \pi \wedge  \delta \varphi \Big),
\end{equation}
does not depend on the radial coordinate; it is a section of $T^*\Sigma_r$. With this observation at hand,  our variational problem for the action \eqref{eq:action} is now well defined. We will ellaborate further in the next section.

\subsection{The constraints}

In order to have a well-defined variational problem, the idea is to use regularization. Specifically, we will use a regularized space $\CM_{r_0}$ with boundary $\partial \CM_{r_0} = \Sigma_{r_0}$ at a large but fixed $r_0$ from the center of the bulk space. If this condition for $r_0$ is satisfied, then this regularized surface $\Sigma_{r_0}$ is diffeomorphic to the boundary $\partial \CM$ at $\infty$. This means that this regulated hypersurface $\Sigma_{r_0}$ can be mapped smoothly to the boundary of the manifold $\CM_{r_0}$. In order for the problem to be well defined in this limit and with the purpose of removing the infinities and ultimately performing holographic renormalization in our model, we consider our action \eqref{eq:action} defined on $\CM_{r_0}$ while adding a generic boundary term $S_b$, that is, we define $S' = S + S_b$.  

Varying $S'$ amounts to
\begin{equation}\label{deltaSprime}
    \begin{aligned}
    \delta S' &= \int_{\CM_{r_0}}  (\text{equations of motion}) + (\CL+\dot{S_b})\Big|_{r_0}\delta r_0 \\ \nonumber
              & + \int_{\Sigma_{r_0}}  \left\{     \left( \pi^{ij}+\frac{\delta S_b}{\delta \gamma_{ij}} \right)\delta \gamma_{ij} +   \left( \pi^{i}+\frac{\delta S_b}{\delta A_{i}} \right)\delta A_i + \left( \pi +\frac{\delta S_b}{\delta \varphi} \right)\delta \varphi \right\}.
    \end{aligned}
\end{equation}

However, the variational problem is not yet well defined. To amend for this at the limit $r_0 \to \infty$ we must further require that
\begin{align}
    \frac{dS'}{dr}\Big|_{r_0} = (L+\dot{S})\Big|_{r_0} \to 0, \qquad  \text{as } r_0 \to \infty,
\end{align}
a requirement that indeed makes the variational problem at infinity well defined for variations of the (induced) bulk fields within the space of generic asymptotic solutions of the equations of motion such that the boundary term $S_b$ at the fixed point $r_0$ is identified with Hamilton's principal function $\CS$ (a function corresponding to solution of the Hamilton-Jacobi equation), which is proportional to the bulk on-shell action, and where the values of the (induced) bulk fields on the regulated hypersurface $\Sigma_{r_0}$ are arbitrary, that is,
\begin{align}
  S_b\Big|_{r_0} = -\CS \Big|_{r_0}.  
\end{align}
Therefore, looking back at the second line of Eq. \eqref{deltaSprime} we realize the precise relation between the conjugate momenta of the induced fields and Hamilton's principal function
\begin{equation}\label{momenta}
    \pi^{ij}\Big|_{r_0} = \frac{\delta \CS}{\delta \gamma_{ij}}\Big|_{r_0}, \quad  \pi^{i}\Big|_{r_0} = \frac{\delta \CS}{\delta A_{i}}\Big|_{r_0},\quad \pi\Big|_{r_0} = \frac{\delta \CS}{\delta \varphi}\Big|_{r_0}.
\end{equation}
Importantly, Hamilton's principal function $\CS$ is identified with the on-shell value of the action \eqref{eq:action} for solutions with arbitrary values for the induced bulk fields on the regulated surface. We can go a step further and make the substitution $\CS \to \CS + S_b$, which amounts to a canonical transformation for the conjugate momenta of the induced bulk fields
\begin{align}
  \Pi^{ij}\Big|_{r_0} = \frac{\delta (\CS + S_b)}{\delta \gamma_{ij}}\Big|_{r_0}, \quad  \Pi^{i}\Big|_{r_0} = \frac{\delta (\CS + S_b)}{\delta A_{i}}\Big|_{r_0},\quad \Pi\Big|_{r_0} = \frac{\delta (\CS + S_b)}{\delta \varphi}\Big|_{r_0}.  
\end{align}

The boundary term, which will later be associated with the divergences of the theory, can be determined by computing the asymptotic form of Hamilton's principal functional $\CS$. This amounts to finding a solution for the Hamilton-Jacobi equation asymptotically, a task that can be achieved by making the following identification. The on-shell action $S$ on the regulated surface $\Sigma_{r_0}$ is identified with $\CS_{r_0} \equiv \CS \Big|_{r_0}$. Then, by taking the radial derivative of $\CS_{r_0}$ we find
\begin{equation}
    \begin{aligned}
    \dot{\CS}_{r_0} &= \frac{\partial \CS_{r_0}}{\partial r_0} + \int_{\Sigma_{r_0}}   \left( \dot{\gamma}_{ij} \frac{\delta}{\delta \gamma_{ij}} + \dot{A}_{i} \frac{\delta}{\delta A_{i}} + \dot{\varphi} \frac{\delta}{\delta \varphi} \right) \CS_{r_{0}} \\
                  &= \frac{\partial \CS_{r_0}}{\partial r_0} +  \int_{\Sigma_{r_0}}   \Big( \dot{\gamma}_{ij} \pi^{ij} + \dot{\gamma}_{i} \pi^{i} + \dot{\varphi} \pi \Big) \\
                  &=  \frac{\partial\CS_{r_0}}{\partial r_0} + H + \CL.
    \end{aligned}
\end{equation}

Note, though, that because $\CS_{r_0}$ is identified with the on-shell action at $\Sigma_{r_0}$ this means that its radial derivative is the Lagrangian, that is, $\dot{\CS}_{r_0} = \CL$, and the previous equation simplifies to 
\begin{align}
    \frac{\partial \CS_{r_0}}{\partial r_0} + H = 0.
\end{align}

Let us remark that here we are essentially dealing with classical supergravity, and, as in any generally covariant theory, the Hamiltonian vanishes identically since it is proportional to the set of constraints we obtained in the previous section. As a consequence, the on-shell action does not depend explicitly on the radial coordinate. The only dependence on the radial coordinate arises indirectly through the dependence of the induced bulk fields on $r$. From the above considerations, we realize that indeed in any generally covariant theory the Hamilton-Jacobi equation amounts to the vanishing of the constraints 
\begin{align}\label{HamJac}
    \mathcal{H} &=0, \\ 
    \mathcal{H}_i &=0, \\
    \mathcal{F} &=0.
\end{align}
For further details, we refer to the excellent lecture notes of Papadimitriou \cite{Papadimitriou2016}.

\section{Recursive solution to the Hamilton-Jacobi equation}

In this section, we will find a recursive solution of the Hamilton-Jacobi equation by following the algorithm presented in \cite{Papadimitriou_2010,Papadimitriou_2011}.

 Once this solution has been obtained, we will be able to determine the boundary term $S_b$ which not only makes the boundary problem well defined at infinity, but will later be identified as the counter-term needed to renormalize the theory. Let us stress the main idea that we will follow here and was first implemented in \cite{Papadimitriou_2011}. 
 
Note that if our variational problem is formulated within a well-defined space of asymptotic solutions, the boundary term will contain no transverse derivatives (to leading order). This means that the full solution of the Hamilton-Jacobi equation admits an expansion in the transverse derivatives. One way to go is to write down an ansatz with all possible and allowed by general covariance terms, containing no transverse derivatives, and substitute it into the Hamilton-Jacobi equation. This ansatz will solve the Hamilton-Jacobi equation to first order and will provide us with an iterative algorithm enabling us to systematically go further to any desired order.

\subsection{The expansion of the solution near $r = \infty$}

Let us begin with Hamilton's principal function $\mathcal{S}_{r}$ given us as the following action functional
\begin{equation}
\mathcal{S}_r = \int_{\Sigma_r}   \mathcal{L}(\gamma, \varphi, A).
\end{equation}
By performing a general variation to both sides above we obtain, via Eqs \eqref{momenta} the following equality
\begin{equation}\label{equality}
\pi^{ij}\delta \gamma_{ij} + \pi_{\varphi}\delta \varphi + \pi^i\delta A_i = \delta \mathcal{L} + \partial_i v^i,
\end{equation}
where $v^i = v^i(\delta \gamma, \delta \varphi, \delta A)$ is some arbitrary vector field. Note that, as $r \to \infty$ the solution of $\CS_r$ admits a derivative expansion of the form
\begin{equation*}\label{expansion}
\CS = \CS_{(0)} + \CS_{(2)} + \CS_{(4)} + \ldots,
\end{equation*}
where we have dropped the radial $r$ index for ease of notation. The leading-order term is given as \cite{Papadimitriou_2011}
\begin{equation}\label{leading}
\CS_{(0)} = \frac{1}{\kappa^2} \int_{\Sigma_r}  \sqrt{-\gamma}\,U(\varphi, A).
\end{equation}

\begin{equation}
\delta_{\gamma} \coloneqq \int   2\gamma_{ij} \frac{\delta}{\delta \gamma_{ij}},
\end{equation}
originally introduced in Ref. \cite{Papadimitriou_2011} and, interestingly, later generalized to non-relativistic holographic theories \cite{Chemissany:2014xpa}. The generalized dilatation operator acts recursively as
\begin{align}\label{recursion}
\delta_\gamma \CS_{(2n)} = (d-2n)\CS_{(2n)},
\end{align}
where $\CS_{(2n)}$ refers to the expansion of Eq. \eqref{expansion}.

For example, consider the action of $\delta_\gamma$ on Eq. \eqref{equality}. For the l.h.s. we obtain

\begin{equation}
    \begin{aligned}
    \delta_{\gamma}(\pi^{ij}\delta \gamma_{ij} + \pi_{\varphi}\delta \varphi + \pi_A^i\delta A_i)&=  \pi^{ij}\delta_{\gamma} \gamma_{ij} + \pi_{\varphi}\delta_{\gamma} \varphi + \pi_A^i\delta_{\gamma} A_i \\
    &= \pi^{ij}\delta_{\gamma} \gamma_{ij} \\
    &= \int   2 \pi^{ij}\gamma_{kl} \, \frac{\delta \gamma_{ij}}{\delta \gamma_{kl}}\\
    & \coloneqq 2\pi_{(2n)}.
    \end{aligned}
\end{equation}
Equivalently, for the r.h.s. we obtain $(d-2n)\mathcal{L}_{(2n)}$ since $\CS_{(2n)} = \int_{\Sigma_r}   \mathcal{L}_{(2n)}$, see \cite{Papadimitriou_2011}. Therefore, we have
\begin{align}
   2\pi_{(2n)} = (d-2n)\CL_{(2n)}. 
\end{align}
The basic idea behind the recursive procedure, the recursive algorithm to be defined shortly, is to identify $\CL_{(2n)}$ at each order.We can insert the leading term $\CS_{(0)}$ into the Hamilton-Jacobi equation to obtain
\begin{equation}\label{equality_triple}
\begin{aligned}
\left(\gamma_{ik}\gamma_{jl} - \frac{1}{d-1}\gamma_{ij}\gamma_{kl} \right)\frac{\delta \CS_{(0)}}{\delta \gamma_{ij}}\frac{\delta \CS_{(0)}}{\delta \gamma_{kl}} &= \frac{-d}{d-1} \frac{\gamma}{4\kappa^2} U^2(\varphi, A) \\
\Rightarrow \,\,\,\,\, \frac{1}{2e^{\alpha \varphi}} \left( \frac{\delta \CS_{(0)}}{\delta A} \right)^2 &= \frac{1}{2e^{\alpha \varphi}} \frac{\gamma}{\kappa^2} \frac{\partial U(\varphi, A)}{\partial A_i} \frac{\partial U(\varphi, A)}{\partial A^i} \\
\Rightarrow \,\,\,\,\, \frac{1}{2}\left( \frac{\delta \CS_{(0)}}{\delta \varphi} \right) &= \frac{1}{2} \frac{\gamma}{2\kappa^2} (\partial_{\varphi} U(\varphi, A))^2, \
\end{aligned}
\end{equation} 
to be left with the following PDE
\begin{equation}
\left(\frac{\partial U}{\partial \varphi}\right)^2 +\frac{1}{e^{\alpha \varphi}} \frac{\partial U}{\partial A_i} \frac{\partial U}{\partial A^i} - \frac{d}{(d-1)}U^2 + V(\varphi) = 0.
\end{equation}
The choice of the potential function $V(\varphi)$ is quite important. For our application at hand, we make the choice $V(\varphi) = e^{b\varphi}$. The choice for this potential is rather important. This is where $Dp$-branes enter the game, as for various $p$ the exponent of the potential takes different values. We have to solve the following equation.
\begin{equation}
\left(\frac{\partial U}{\partial \varphi}\right)^2 +\frac{1}{e^{\alpha \varphi}} \frac{\partial U}{\partial A_i} \frac{\partial U}{\partial A^i} - \frac{d}{(d-1)}U^2 +  ce^{b\varphi} = 0.
\end{equation}
Note that $U$ cannot depend on the gauge field $A$ since it is impossible to construct a Lorentz invariant combination of the gauge fields alone without breaking gauge invariance.  As a result, we are only left with the following, much simpler, differential equation to solve.
\begin{equation}\label{eq:pde}
\left(\frac{\partial U}{\partial \varphi}\right)^2 - \frac{d}{(d-1)}U^2 +  ce^{b\varphi} = 0.
\end{equation}
Furthermore, any functional $f$ for which \eqref{equality_triple} is valid will suffice to determine the leading term of \eqref{leading}. Let us set $a = \frac{d}{(d-1)}$. 
Then, Eq. \eqref{eq:pde} takes the form
\begin{equation}
 U'(\varphi)^2 - aU(\varphi)^2 +  ce^{b\varphi} = 0,
\end{equation}
and we can try an ansatz of the form $U(\varphi) = ge^{h\varphi}$. Eq. \eqref{eq:pde} becomes
\begin{eqnarray}
h^2g^2e^{2h\varphi} - ag^2e^{2h\varphi} + ce^{b\varphi} &=& g^2(h^2 - a)e^{2h\varphi}  + ce^{b\varphi} \\
														&=& 0,
\end{eqnarray}
by requiring that $h = b/2$ and $g = \pm \sqrt{\frac{c}{a^2 - b^2/4}} $, we get the solution
\begin{equation}\label{solution_U}
U(\varphi) = \pm \sqrt{\frac{c}{a^2 - \, b^2/4}}  e^{\frac{b}{2} \varphi}.
\end{equation}
This solution can now be substituted into $\CS_{(0)}$ and we can proceed recursively, via Eq. \eqref{recursion}, to obtain the next term in the expansion $\CS =\CS_{(0)} + \CS_{(2)} + \ldots$, as we discuss in the next subsection.

\subsection{The recursive solution $\CR_{(2n)}$}

In this subsection we dive deeper into the recursive algorithm \cite{Papadimitriou_2011}.
Inserting the solution \eqref{solution_U} into the leading term in the expansion, $\CS_{(0)}$ allows us to obtain the following result. 
\begin{equation}
\CS_{(0)} = \frac{1}{\kappa^2} \int_{\Sigma_r}   \sqrt{-\gamma} \, ge^{\frac{b}{2} \varphi}.
\end{equation}
To get higher-order solutions, we proceed in the following way. We insert the expansion \eqref{expansion} into the Hamilton-Jacobi equation $\mathcal{H}$ and then we aim to match terms of equal $\delta_{\gamma}$ eigenvalue. Explicitly, we have
\begin{equation}
    \begin{aligned}
    \mathcal{H} &= -\frac{2\kappa^2}{\sqrt{-\gamma}}  \Bigg( \frac{1}{2} \sum_{n=0}^{\infty}\sum_{m=0}^{\infty} \left( \frac{\partial \CS_{(n)}}{\partial \varphi} \frac{\partial \CS_{(m)}}{\partial \varphi} + \frac{1}{2e^{\alpha \varphi}} \frac{\partial \CS_{(n)}}{\partial A_i} \frac{\partial \CS_{(m)}}{\partial A^i} \right) \\
    & \quad + \left( \gamma_{ik} \gamma_{jl} - \frac{1}{d-1} \gamma_{ij}\gamma_{kl} \right) \sum_{n=0}^{\infty}\sum_{m=0}^{\infty}\frac{\partial \CS_{(n)}}{\partial \gamma_{ij}} \frac{\partial \CS_{(m)}}{\partial \gamma_{kl}}    \Bigg) \\
& \quad +  \frac{\sqrt{-\gamma}}{2\kappa^2} \left( -R[\gamma] + \frac{1}{2}  \gamma^{ij} \partial_i \varphi \partial_j \varphi + V(\varphi) + \frac{e^{\alpha \varphi}}{4}F_{ij}F^{ij}\right). \
    \end{aligned}
\end{equation}
The above sum satisfies the following Cauchy series property.
\begin{equation}
\sum_{n=0}^{\infty}\sum_{m=0}^{\infty} a_n  b_m = \sum_{n=}^{\infty} \sum_{m=0}^{n} a_m b_{n-m} 
\end{equation}
Using this property, we can match terms that have the same weight under the $\delta_{\gamma}$ operator to obtain the following. 
\begin{equation}\label{Uprime}
U'(\varphi)\frac{\delta}{\delta \varphi} \int \CL_{(2n)} - \frac{d-2n}{d-1}U(\varphi)\CL_{(2)} = \CR_{(2n)}.
\end{equation}
In the lowest order, the identification reads 
\begin{align}\label{R2}
    \CR_{(2)} =  -\frac{\sqrt{-\gamma}}{2\kappa^2} \left( -R[\gamma] + \frac{1}{2} \partial_i \varphi \partial^i \varphi \right),
\end{align}
with the generic prescription given by
\begin{equation}\label{recR}
\begin{aligned}
\mathcal{R}_{(2n)}&=-\frac{2\kappa^2}{\sqrt{-\gamma}} \sum_{m=1}^{n-1} \Big(  \pi^{ij}_{(2m)} \pi_{(2(n-m)) \,\,ij}- \frac{1}{d-1} \pi_{(2m)} \pi_{(2(n-m))} \\
                      &\quad + \frac{1}{2}\pi_{(2m)}^{\varphi} \pi_{(2(n-m))}  + \frac{1}{2e^{\alpha \varphi}} \pi_{(2m)}^i \pi_{(2(n-m))\,\, i} \Big), \qquad n>1. \
\end{aligned}
\end{equation}
Although we have not explicitly described the recursive algorithm yet, we will explain in the next subsection that it involves Eqs. \eqref{Uprime} and \eqref{recR}. The first one is a linear inhomogeneous pde involving derivatives with respect to the dilaton only. The observation here is that every time we solve the above equation we get a further correction to $\CS$ and therefore to the boundary term $S_b$.

\subsection{Computing the solution $\CL_{(2n)}$}

The relation between $\CL_{(2n)}$ and $\CR_{(2n)}$, becomes clearer by rewriting Eq. \eqref{Uprime} as
\begin{equation}
\frac{\delta}{\delta \varphi} \int \CL_{(2n)} - \frac{d-2n}{d-1} \frac{U(\varphi)}{U'(\varphi)}\CL_{(2)} = \frac{\CR_{(2n)}}{U'(\varphi)},
\end{equation}
which has the generic form of 
\begin{equation} \nonumber
\frac{d F(\varphi)}{d \varphi} - c\frac{G(\varphi)}{G'(\varphi)}F(\varphi) = \frac{H(\varphi)}{G'(\varphi)}.
\end{equation}
Eq. \eqref{Uprime}, admits a homogeneous solution of the form
\begin{equation}\label{solution2}
\CL_{(2n)}^{\text{hom.}} = \mathcal{F}_{(2n)}[\gamma, A] \, e^{-(d-2n)\bar{A}(\varphi)}.
\end{equation}
Here, we used a function $\bar{A}$ that will be proven to be very useful later on. Specifically, its definition is
\begin{equation}
\bar{A}(\varphi) \coloneqq -\frac{1}{d-1} \int^{\varphi} d\bar{\varphi} \frac{U(\bar{\varphi})}{U'(\bar{\varphi})}
\end{equation}
where $\mathcal{F}_{(2n)}$ is a covariant function of the induced metric $\gamma$ and the gauge field $A$. The solution \eqref{solution2} can easily be verified by direct substitution. However, the inhomogeneous solution of Eq. \eqref{Uprime} is more interesting. The most general solution for this equation takes the form
\begin{equation}
\CL_{(2n)}^{\text{inhom.}} = e^{-(d-2n)\bar{A}(\varphi)}\mathcal{F}_{(2n)}[\gamma, A, \varphi], 
\end{equation}
where
\begin{equation}
\CF_{(2n)} = \int^{\varphi} \frac{d \bar{\varphi}}{U'(\varphi)} e^{(d-2n)\bar{A}(\varphi)} \CR_{(2n)}(\bar{\varphi}).
\end{equation}
Varying the inhomogeneous solution 
\begin{equation}\label{solution_inhom}
\begin{aligned}
\delta_{\varphi} \CL_{(2n)}^{\text{inhom.}}  = -(d-2n)\bar{A}'e^{(d-2n)\bar{A}(\varphi)}(\varphi) \CF_{(2n)} + e^{-(d-2)\bar{A}(\varphi)} \delta_{\varphi} \CF_{(2)} ,
\end{aligned}
\end{equation}
we find that the function $\CF_{(2n)} $ satisfies
\begin{equation}\label{another_eq}
\frac{\delta \varphi}{U'(\varphi)} e^{a_n} \CR_{(2n)} = \delta_{\varphi}\CF_{(2n)} + e^{a_n} D_i v^i(\varphi, \delta \varphi),
\end{equation}
for some arbitrary vector field $v^i(\varphi, \delta \varphi)$ due to the fact that the inhomogeneous solution \eqref{solution_inhom} is defined up to such a vector field. Here, we have defined another useful quantity, $a_n \coloneqq -(d-2n)\bar{A}(\varphi)$.

This integration formula provides an algorithmic procedure to iteratively evaluate Hamilton's principal function $\CS$. This can be done by obtaining $\CL_{(2n)}$ (up to a noncontributing total derivative term) from the same order source term $\CR_{(2n)}$. Then, by differentiating $\CL_{(2n)}$ with respect to the induced fields, we obtain the conjugate momenta of the corresponding order. Once we have the conjugate momenta, we can iteratively compute the next order source term $\CL_{(2(n+1))}$ to the desired order.

In the next subsection we will carry out this algorithm up to $n=1$ which is sufficient for the cases of systems involving $D1$-branes and $D2$-branes. However, before proceeding, let us note that generally, at each order $n$ the source term $\CR_{(2n)}$ and the corresponding order inhomogeneous solution of Eq. \eqref{Uprime}  can be written as a sum of (to be specified) tensors as
\begin{equation*} 
\CR_{(2n)} = -\frac{\sqrt{-\gamma}}{2\kappa^2} \, \sum_{I=1}^{N_n} c_n^I(\varphi) \mathcal{T}_n^I, 
\end{equation*}
in analogy to Eq. \eqref{R2}. Additionally, the Lagrangian $\CL$, and therefore Hamilton's principal function $\CF$, can be written as the product of certain functions involving such tensors
\begin{equation*}  
\CL_{(2n)} = -\frac{\sqrt{-\gamma}}{2\kappa^2} \, \sum_{I=1}^{N_n} \mathcal{P}_n^I(\varphi) \mathcal{T}_n^I. 
\end{equation*}
Here, $c_n^I(\varphi)$ and $ \mathcal{P}_n^I(\varphi)$ are scalar functions of $\varphi$ and $ \mathcal{T}_n^I$ are tensor fields that involve fields other than $\varphi$ and possibly derivatives of $\varphi$. From the last equation, we are in principle able to read the counter-term action since we know how it is related to $\CL_{(2n)}$. For more details, see \cite{Papadimitriou_2011}.

\subsection{Leading order contribution of $\CL_{(2n)}$}

In the simplest case, that is, when $n=1$, we compute $\CL_{(2)}$. Once this is done, the computation of $\CL_{(4)}$, or any desired $\CL_{(2n)}$, is relatively straightforward algorithmically. However, for the higher-order terms we will have to perform a non-trivial integration when $\CR_{(2n)}$ involves derivatives of $\varphi$. In the leading-order computation, $\CR_{(2)}$ involves a trivial term corresponding to the Ricci scalar and a term of the form  
\begin{equation} 
\CR_{(2n)}(\varphi) = r_{1^m}(\varphi)t^{i_1\ldots i_m} \partial_{i_1}\varphi \ldots \partial_{i_m} \varphi, 
\end{equation}
For such term we propose that $\CF_{(2n)}$ has the form
\begin{equation}\label{F2n}
\CF_{(2n)} = e^{a_n} \Big( \alpha(\varphi) t^{i_1 \ldots i_m} \partial_{i_1}\varphi \ldots \partial_{i_m} \varphi + D_i ( \beta(\varphi) t^{ i_1 \ldots i_m} \partial_{i_1}\varphi \ldots \partial_{i_m}\varphi ) \Big).
\end{equation}
To write the complete form of the inhomogeneous solution \eqref{solution_inhom} we have to determine the coefficients $\alpha(\varphi)$ and $\beta(\varphi)$ which amounts to varying \eqref{F2n} and substituting into \eqref{another_eq}. Explicitly, the variation yields
\begin{equation}
\begin{aligned}
\delta_{\varphi} \CF_{(2n)} &= e^{a_n}(a'_n\alpha+ \alpha' - a_n''\beta) t^{i_1 \ldots i_m} \partial_{i_1} \varphi \ldots \partial_{i_m} \varphi \delta \varphi + e^{a_n} D_i \tilde{u}^i \\
&\quad + e^{a_n} m \alpha t^{i_1 \ldots i_m} \partial_{i_1} \delta \varphi \ldots \partial_{i_m} \varphi - e^{a_n} a_n \beta \partial_{i_1} \delta \varphi \ldots \partial_{i_m} \varphi,
\end{aligned}
\end{equation}
while substituting into \eqref{another_eq} yields 
\begin{eqnarray}
\beta = \frac{m}{a'_n}\alpha \\
 a'_n\alpha + \alpha' - m\frac{a''_n}{a'_n}\alpha = \frac{r_{1^m}}{U'}.
\end{eqnarray}
However, $\beta(\varphi)$ only contributes to the total derivative term, thus we are only interested in $\alpha(\varphi)$. The latter of the two resembles a chain rule derivative;  we can multiply both sides by ${a'}_n^{-m}$ to get
\begin{equation}
 {a'}_n^{-m}\alpha' + a'_n {a'}_n^{-m} \alpha - m {a'}_n^{-m-1} {a''}_n \alpha = {a'}_n^{-m} \frac{r_{1^m}}{U'},
\end{equation} 
which can be rewritten as
\begin{equation}
\frac{d}{d\varphi} ( \alpha {a'}_n^{-m} e^{a_n}) = e^{a_n}  {a'}_n^{-m} \frac{r_{1^m}}{U'}.
\end{equation}
This provides us with a solution for $\alpha(\varphi)$ that takes the form
\begin{equation} 
\alpha(\varphi) = e^{-a_n} {a'}_n^{m} \int^{\varphi} \frac{d\bar{\varphi}}{U'} e^{a_n} {a'}_n^{-m} r_{1^m}(\bar{\varphi}) = \fint_{n,m}^{\varphi} r_{1^m}(\bar{\varphi}),
\end{equation}
with the definition 
\begin{equation*}
\fint_{n,m}^{\varphi} \coloneqq e^{-a_n} {a'}_n^{m} \int^{\varphi} \frac{d\bar{\varphi}}{U'} e^{a_n} {a'}_n^{-m}.
\end{equation*}
For our problem, we have  
\begin{eqnarray*}
m=2, && r_{1^0} = -1 \\
t^{i_1 i_2} = \gamma^{ij}, && r_{1^2} = \frac{1}{2}, \
\end{eqnarray*}
and in this case the coefficient $\alpha$ takes the form 
\begin{equation} 
\alpha(\varphi) = e^{-a_1} {a'}_1^{2} \int^{\varphi} \frac{d\bar{\varphi}}{U'} e^{a_1} {a'}_1^{-2}  (r_{1^0} + r_{1^2}) = \fint_{1,(1,2)}^{\varphi}  (r_{1^0} + r_{1^2})  .
\end{equation}
where $a_1 = -(d-2)\bar{A}(\varphi)$. The case of interest is the one where $U(\varphi)$ is given by Eq. \eqref{solution_U}.  Then $\bar{A}(\varphi)$ is $- \frac{2}{b} \frac{1}{d-1}\varphi$ and 
\begin{equation*}
a_1 =\frac{2}{b} \frac{d-2}{d-1}\varphi,
\end{equation*}
and 
\begin{equation*}
a'_1 =\frac{2}{b} \frac{d-2}{d-1}. 
\end{equation*}
We are now able to determine $\CL_{(2)}$ completely using the results derived above. Recall that the source term at order $n=1$ is given by Eq. \eqref{R2}.
Thus, for $n=1$, we obtain Table \ref{table1}
\begin{table}[!tbh]
\begin{center}
\begin{tabular}{| l  c c c | } \Xhline{5\arrayrulewidth}
  $I$  & $c_1^I(\varphi)$    & $\mathcal{T}_1^I(\varphi)$ &$\CP_{1}^I(\varphi)$  \\ \hline
  $1$ & $-1$& $R$ & $-2\Xi(\varphi)$ \\
  $2$ & $\frac{1}{2}$  & $\partial_i \varphi \partial^i \varphi$ & $-M(\varphi)$ \\ \Xhline{3\arrayrulewidth}
\end{tabular}
\end{center}
\caption{The coefficients and tensors appearing at the $n=1$ source term.}
\label{table1}
\end{table}
The functions $\Xi(\varphi)$ and $M(\varphi)$ are determined by 
\begin{equation}
\begin{aligned}
\CP_1^1(\varphi) &= \fint_{1,1}^{\varphi} r_{1^1} \\
              &= -\frac{e^{- \frac{b}{2}\varphi  }}{g\left( \frac{d-2}{d-1} - (\frac{b}{2})^2 \right)}   \\
              &\equiv -2 \Xi(\varphi), \
\end{aligned}
\end{equation}
while
\begin{equation}
\begin{aligned}
\CP_1^2(\varphi) &= \fint_{1,2}^{\varphi} r_{1^2} \\
              &= \frac{1}{2} \frac{e^{- \frac{b}{2}\varphi  }}{g\left( \frac{d-2}{d-1} - (\frac{b}{2})^2 \right)} \\
              &\coloneqq - M(\varphi). \
\end{aligned}
\end{equation}

Now we indeed have everything we need to determine the inhomogeneous order $n=1$ solution which is
\begin{equation}
\CL_{(2)} = -\frac{\sqrt{-\gamma}}{2\kappa^2} \left( -2\Xi(\varphi)R -M(\varphi) \partial_i \varphi \partial^i \varphi \right).
\end{equation}
This formula concludes the computation of the $n=1$ Lagrangian and therefore the corresponding solution of the Hamilton-Jacobi equation. Combining $\CL_{(0)}$ and $\CL_{(2)}$ we obtain the required counter-term action for the brane system at hand. Concluding this section, let us summarize the iterative algorithm that we have used and how we can go further.

We begin with Eq. \eqref{Uprime} or Eq. \eqref{another_eq} for the $2n$-order solution. We solve the inhomogeneous equation and obtain the solution $\CL_{(2n)}$. Next, we use this solution to find the corresponding conjugate momenta and then compute $\CR_{(2n+2)}$. In turn, using the corresponding $\fint$ we compute $\CL_{(2+2)}$. The iterative procedure continues by finding the corresponding conjugate momenta, etc. The iterative algorithm is schematically described in Table \ref{table2}.

\begin{table}[h]
\begin{center}
\begin{tabular}{ c  c c c  c } 
$\CR_{(2n)}$ & $\overset{\fint}{\longrightarrow}$  & $\CL_{(2n)}$ & $\overset{\delta}{\longrightarrow}$ & $\{ \pi_{(2n)} \}$ \\
&&&& $	\Big\downarrow$ \\
$\{ \pi_{(2n+2)} \}$ & $\overset{\delta}{\longleftarrow}$  & $\CL_{(2n+2)}$ & $\overset{\fint}{\longleftarrow}$ & $\CR_{(2n+2)}$ \\
$\Big\downarrow$&&&& 	 \\
$\CR_{(2n+4)}$ & $\overset{\fint}{\longrightarrow}$  & $\ldots$ &  & \\
\end{tabular}
\end{center}
\label{table2}
\caption{A schematic representation of the iterative algorithm we use to determine Hamilton's principal function.}
\end{table}

Having obtained $\CL_{(2)}$ we have everything we need in order to determine the boundary term $S_b$ for the systems of $D1$-branes and $D2$-branes. Holographic renormalization and the holographic dictionary for $D3$-branes well known. For $D4$-branes our method works as well but it is required to go one order beyond, to $\CL_{(4)}$. For $Dp$-branes with $p>4$ we will find that there are various subtleties in using the algorithm specified above and their treatment might require more subtle methods.

\section{The generalized Fefferman-Graham expansions}
In the previous section, we managed to solve the Hamilton-Jacobi equation in an asymptotic way. Our next task is to construct the space of asymptotic solutions for the bulk-induced fields. We will fix the shift function to $N^i=1$ and the lapse function $N=0$ and find the asymptotic solutions via the Fefferman-Graham expansion \cite{Fefferman95}. For that we need to recall the first-order flow equations we found explicitly and can be written as follows in terms of the Hamilton principal function \cite{Papadimitriou_2010, Papadimitriou_2011, Papadimitriou2018} 
\begin{eqnarray}
\dot{\gamma}_{ij} &=&\frac{4\kappa^2}{\sqrt{-\gamma}} (\gamma_{ik}\gamma_{jl} - \frac{1}{d-1} \gamma_{ij}\gamma_{kl} ) \frac{\delta \mathcal{S}}{\delta \gamma_{kl}}, \\
\dot{\varphi} &=&  \frac{2\kappa^2}{\sqrt{-\gamma}} \frac{\delta \mathcal{S}}{\delta \varphi }, \\
\dot{A}_i &=& \frac{2\kappa^{2}}{\sqrt{-\gamma}} e^{-\alpha \varphi}   \frac{\delta \mathcal{S}}{\delta A_i }. \
\end{eqnarray}
Note that we have also gauge fixed $a=0$. Using the results of the previous subsection, the computation of the conjugate momenta yield
\begin{equation}
\begin{aligned}
    \pi_{(2)}^{ij} &= -\frac{\sqrt{-\gamma}}{\kappa^2} \Bigg( \Xi R^{ij} - \Xi' D^i D^j \varphi + \frac{1}{2} ( M - 2\Xi'' ) \partial^i\varphi \partial^j \varphi  \\
                 & -\frac{1}{2} \gamma^{ij} \left( \Xi R - \Xi' \Box_{\gamma} \varphi + \frac{1}{2} (M - 4\Xi'') \partial_k \varphi \partial^k \varphi  \right)     \Bigg),  \\  
            & \\
\pi_{(2)} &= -\frac{\sqrt{-\gamma}}{\kappa^2} (- \Xi' R - M'\partial_i \varphi \partial^i \varphi + M \Box \varphi), \\  
            & \\
\pi_{(2)}^i &= 0.
\end{aligned}
\end{equation}

\subsection{Asymptotic expansion of the $n=0$ flow equations}

We begin our analysis with the dilaton $\varphi$. Using $S_{(0)}$ we get the 0-th order flow equations. Recall that 
\begin{equation*}
\CS_{(0)} = \frac{1}{\kappa^2} \int_{\Sigma_r} \sqrt{-\gamma} \, ge^{\frac{b}{2}\varphi},
\end{equation*}
and varying with respect to $\varphi$ we get
\begin{align}
    \frac{\delta S_{(0)}}{\delta \varphi} = \frac{\sqrt{-\gamma}}{\kappa^2} g\frac{b}{2}e^{\frac{b}{2}\varphi}.
\end{align}
Therefore, at zero-order
\begin{equation}
\overset{{(0)}}{\dot{\varphi}} = g\frac{b}{2}e^{\frac{b}{2}\overset{(0)}{\varphi}},
\end{equation}
and the solution reads
\begin{equation}\label{zeroordersolution}
\overset{(0)}{\varphi}(r,x) = - \frac{2}{b} \log \left(\left| - g\frac{b^2}{2} r +  \varphi{(0)}(x) \right| \right) .
\end{equation}
The $(0)$ overscript indicates the order of the Fefferman-Graham expansion to which it belongs, while $ \varphi_{(0)}(x)$ is an integration constant. Let us make an observation for the induced metric. It has been shown in \cite{Fefferman95} that near the boundary the induced metric can be written as follows
\begin{equation}\label{metricexp}
\gamma_{ij}(r,x) = e^{2\bar{A}(r)}\Big ( \bar{g}_{(0)ij}(x) + \ldots \Big).
\end{equation}
By varyinf $\CS_{(0)}$ with respect to the induced metric, we find
\begin{equation}
\frac{ \delta \CS_{(0)}}{ \delta \gamma_{ij}} = - \frac{1}{2} \frac{\sqrt{-\gamma}}{2\kappa^2} \gamma^{ij}U(\varphi), 
\end{equation}
and using \eqref{metricexp} we arrive at
\begin{equation}
\begin{aligned}
    \dot{ {\overset{(0)}{\gamma} }}_{ij} &= e^{2\bar{A}(r)} \Big( \bar{g}_{(0)i}^{l} \bar{g}_{(0)jl} - \frac{1}{d-1} \bar{g}_{(0)ij} \bar{g}_{(0)l}^{l} \Big)U(\varphi) \\
                    &= -\frac{2}{d-1} e^{2\bar{A}(r)} \bar{g}_{(0)ij} U(\varphi) \\ 
                    &=  -\frac{2g}{d-1}  \bar{g}_{(0)ij} \,e^{2\bar{A}(r)} e^{\frac{b}{2}\varphi(r)} \\ 
                    &= -\frac{2g}{d-1}  \bar{g}_{(0)ij} \left( -\frac{1}{2}gb^2r + \varphi_{(0)} \right)^{\alpha-1},
\end{aligned}
\end{equation}
where $\alpha = \frac{4}{b^2(d-1)}$ is a useful constant.
Next, we focus on $\bar{A}$ whereby using \eqref{metricexp} we can rewrite it as a function of $r$ only. 
\begin{equation}
\bar{A}(r) = \frac{2}{b^2}\frac{1}{d-1} \log \left( -g\frac{b^2}{2}r + \varphi_{(0)} \right).
\end{equation}
Therefore, we have
\begin{equation}
\dot{ {\overset{(0)}{\gamma} }}_{ij} =  \frac{g}{d-1}  \bar{g}_{(0)ij} \left( -\frac{1}{2}gb^2r + \varphi_{(0)} \right)^{\alpha-1},
\end{equation}
and the solution is
\begin{equation*} 
{\overset{(0)}{\gamma} }_{ij}(r,x) = -\frac{4}{ab^2(d-1)} \bar{g}_{(0)ij}  \left( -\frac{1}{2}gb^2r + \varphi_{(0)}(x) \right)^{\alpha},
\end{equation*}
or
\begin{equation}\label{1416}
{\overset{(0)}{\gamma} }_{ij}(r,x) = \frac{1}{2} \bar{g}_{(0)ij}  \left( -\frac{1}{2}gb^2r + \varphi_{(0)}(x) \right)^{\alpha} .
\end{equation}
Performing a similar, but easier, analysis for the gauge field, we find  
\begin{equation}
{\overset{(0)}{A} }_{i}(r,x) = 0,
\end{equation}
thus, the solution is given by 
\begin{equation}
{\overset{(0)}{A} }_{i}(r,x) = {\overset{(0)}{A} }_{i}(x).
\end{equation}
It turns out that we need not further worry about the gauge field. The field strength operator we find in the action of our problem has weight 4 under the generalized dilatation operator. This means that it enters into the asymptotics on the order $n=2$ for which we will not perform the analysis here. However, let us make a remark about $\delta_{\gamma}$. What this operator actually does is count the number of inverse metrics appearing in a given operator. The term $F_{\mu \nu}F^{\mu \nu}$ involves two inverse metrics as opposed to, say, $R = R_{\mu \nu} g^{\mu \nu}$. The number of inverse metrics plays a role in the asymptotics of the induced fields, and more inverse metrics contribute in higher-order terms. Finally, let us remark that the leading terms of the asymptotic expansions of the induced bulk fields $\bar{g}_{(0)ij}, \varphi$, are identified as the sources of the dual operators on the dual field theory side.

\subsection{Asymptotic expansion of the order $n=1$ flow equations}
Let us proceed and compute the next terms of the asymptotic expansions, subleading terms. Now we begin with the induced metric and we will use the $n=1$ conjugate momenta and the flow equation for the induced metric. We have
\begin{eqnarray}\label{gammaeq} 
\dot{\gamma}_{ij} &=& -\frac{4\kappa^2}{\sqrt{-\gamma}} \left( \gamma{ik}\gamma{jl} - \frac{1}{d-1}\gamma{ij}\gamma{kl} \right) \frac{\delta}{\delta{kl}}( \CS_{(0)} + \CS_{(2)} + \ldots  ), \
\end{eqnarray}
and now we can make a simplification by defining $\tilde{\pi}^{ij} = \frac{1}{\sqrt{-\gamma}} \pi^{kl} $, where, using earlier results, we have
\begin{equation}
\begin{aligned}
    \tilde{\pi}_{(0)}^{ij} &= \frac{1}{\kappa^2} ge^{\frac{b}{2}\varphi} \, \gamma^{ij} \\
\tilde{\pi}_{(2)}^{ij} &= -\frac{1}{\kappa^2}  \Bigg( \Xi R^{ij} - \Xi' D^i D^j \varphi + \frac{1}{2} ( M - 2\Xi'' ) \partial^i\varphi \partial^j \varphi  \\
                 & -\frac{1}{2} \gamma^{ij} \left( \Xi R - \Xi' \Box_{\gamma} \varphi + \frac{1}{2} (M - 4\Xi'') \partial_k \varphi \partial^k \varphi  \right)     \Bigg).  \
\end{aligned}
\end{equation}
Using Eq. \eqref{gammaeq} we can find the following quantity of interest, where the order 2 terms are not only given, as we naively would think by $\tilde{\pi}_{(2)}$ contributions, rather it is a bit more involved. We have to make every combination allowed as following
\begin{equation}
\begin{aligned}
    \dot{ {\overset{(2)}{\gamma} }}_{ij} &= -4ge^{\frac{b}{2}\varphi} \Bigg\{ \left( \OGG_{ik}\OG_{jl} + \OG_{jk} \OGG_{jl} - \frac{1}{d-1} \OGG_{ij} \OG_{kl} - \frac{1}{d-1} \OG_{ij} \OGG_{kl}   \right) \OG^{kl}   \\
                                             &\quad - \left( \OG \OG_{jl} - \frac{1}{d-1}\OG_{ij} \OG_{kl}  \right) \OGG^{kl} \Bigg\}  - 4\kappa^2 \left(  \OG_{ik} \OG_{jl} - \frac{1}{d-1}\OG_{ij} \OG_{kl}  \right)\tilde{\pi}_{(2)}^{kl}. \
\end{aligned}
\end{equation}
Algebra manipulations lead to an inhomogeneous pde for the radial coordinate $r$ as follows
\begin{equation}
\dot{ {\overset{(2)}{\gamma} }}_{ij} - \frac{4g}{d-1}e^{\frac{b}{2}\varphi}  \OGG_{ij}=  - 4\kappa^2 \left(  \OG_{ik} \OG_{jl} - \frac{1}{d-1}\OG_{ij} \OG_{kl}  \right)\tilde{\pi}_{(2)}^{kl}.  
\end{equation}
The homogeneous solution of the previous pde is given by 
\begin{equation}
{\OGG}_{ij}^{\text{hom.}} = {\lambda}_{ij}(x) \exp \left[{-\frac{4g}{d-1}\int^r dr' \, e^{\frac{b}{2}\varphi(r)}}\right],
\end{equation}
and using the 0-order solution \eqref{zeroordersolution} we find 
\begin{equation}
{\OGG}_{ij}^{\text{hom.}} =  {\lambda}_{ij}(x)  e^{Q(r)},
\end{equation}
where 
\begin{equation} 
Q(r) =   \frac{8}{b^2(d-1)} \log \left(-g\frac{b^2}{2}r + \varphi{(0)} \right) = 2\alpha \log \left(-g\frac{b^2}{2}r + \varphi_{(0)} \right) . 
\end{equation}
The inhomogeneous solution will be given by 
\begin{equation}
\begin{aligned}
    {\OGG}_{ij}^{\text{inhom.}} &= e^{Q(r)} \int^r dr' e^{-Q(r')} \left(-4\kappa^2 \left(  \OG \OG_{jl} - \frac{1}{d-1}\OG_{ij} \OG_{kl}  \right)\tilde{\pi}_{(2)}^{kl} \right)  \\ 
		 & \\ 
                 &=4\left( \bar{g}_{(0)ik}\bar{g}_{(0)jl} - \frac{1}{d-1}\bar{g}_{(0)ij}\bar{g}_{(0)kl} \right)\, e^{Q(r)} \times \\ 
                 & \,\,\,\,\,\,\,\,\,\,\,\,\,\,\,\,\, \int^r dr' \, e^{-Q(r)} \Bigg\{   \Xi R^{kl}[\bar{g}_{(0)}] - \Xi' D_{(0)}^k D_{(0)}^l \varphi + \frac{1}{2} ( M - 2\Xi'' ) D_{(0)}^k\varphi D_{(0)}^l \varphi  \\
                 &  \,\,\,\,\,\,\,\,\,\,\,\,\,\,\,\,\, -\frac{1}{2} \bar{g}_{(0)}^{kl} \left( \Xi R[\bar{g}_{(0)}] - \Xi' \Box_{\bar{g}_{(0)}} \varphi + \frac{1}{2} (M - 4\Xi'') D_{(0)k} \varphi D_{(0)}^k \varphi  \right)   \Bigg\}, \
\end{aligned}    
\end{equation}
keeping in mind that
\begin{equation*}
e^{Q(r)} =  \left(-g\frac{b^2}{2}r + \varphi_{(0)} \right)^{2\alpha}, \qquad e^{-Q(r)} =  \left(-g\frac{b^2}{2}r + \varphi_{(0)} \right)^{-2\alpha}.
\end{equation*}
Now we can substitute the explicit expressions for $\Xi(\varphi), M(\varphi)$, which we found earlier, as well as their derivatives, into the previous equation to obtain
\begin{equation}
\begin{aligned}
{\OGG}_{ij}^{\text{inhom.}} &=4\left( \bar{g}_{(0)ik}\bar{g}_{(0)jl} - \frac{1}{d-1}\bar{g}_{(0)ij}\bar{g}_{(0)kl} \right)\, e^{Q(r)}   \\  
                 & \quad \times \int^r dr' \, e^{-Q(r)} \Bigg\{   \frac{1}{z}e^{-\frac{b}{2}\varphi} R^{kl} + \frac{b}{2z}e^{-\frac{b}{2}\varphi} D^k D^l \varphi - \frac{b^2+1}{4z}e^{-\frac{b}{2}\varphi} \partial^k\varphi \partial^l \varphi  \\
                 &  \quad - \bar{g}_{(0)}^{kl} \left( \frac{1}{2z}e^{-\frac{b}{2}\varphi} R + \frac{b}{4z}e^{-\frac{b}{2}\varphi} \Box_{\bar{g}_{(0)}} \varphi + \frac{2b^2-1}{8z}e^{-\frac{b}{2}\varphi} \partial_k \varphi \partial^k \varphi  \right)   \Bigg\}, \
\end{aligned}
\end{equation}
where we have defined the constant
\begin{equation} 
z= g\left( \frac{d-2}{d-1} - \Big( \frac{b}{2} \Big)^2 \right).
\end{equation}
Therefore, we end up with 
\begin{equation}
\begin{aligned}
{\OGG}_{ij}^{\text{inhom.}} &=\frac{4}{z}\left( \bar{g}_{(0)ik}\bar{g}_{(0)jl} - \frac{1}{d-1}\bar{g}_{(0)ij}\bar{g}_{(0)kl} \right)\, \left(-g\frac{b^2}{2}r + \varphi_{(0)} \right)^{2\alpha} \times \\ 
                 &\quad \int^r dr' \,  \left(-g\frac{b^2}{2}r + \varphi_{(0)} \right)^{-2\alpha}  \left(-\frac{1}{4}gb^2r + \varphi_{(0)} \right) \Bigg\{    R^{kl}[\bar{g}_{(0)}] + \frac{b}{2} \Bigg( -\frac{2 D_{(0)}^{k}D_{(0)}^{l} \varphi_{(0)}    }{b\left( -\frac{1}{4}gb^2r + \varphi_{(0)} \right)}  \\ 
                 &\quad -\frac{2 D_{(0)}^{k} \varphi_{(0)} D_{(0)}^{l} \varphi_{(0)}    }{b^2\left( -\frac{1}{4}gb^2r + \varphi_{(0)} \right)^2}  \Bigg) - \frac{b^2+1}{4} \left( 4 \frac{ D_{(0)}^{k}\varphi_{(0)} D_{(0)}^{l} \varphi_{(0)}   }{ b^2\left( -\frac{1}{4}gb^2r + \varphi_{(0)} \right)^2   } \right)  \\ 
                 &  \quad - \bar{g}_{(0)}^{kl} \Bigg( \frac{1}{2} R[\bar{g}_{(0)}] + \frac{b}{4} \left( \frac{2D_{(0)}^{m} \varphi_{(0)} D_{(0)m} \varphi_{(0)}  }{ b\left( -\frac{1}{4}gb^2r + \varphi_{(0)}  \right)^2 }   - \frac{2 \Box_{\bar{g}_{(0)}} \varphi_{(0)} }{b\left( -\frac{1}{4}gb^2r + \varphi_{(0)} \right)  }   \right) \\
                 &  \quad+ \frac{2b^2-1}{8} \left( \frac{4D_{(0)}^{m} \varphi_{(0)} D_{(0)m} \varphi_{(0)}  }{b^2\left( -\frac{1}{4}gb^2r + \varphi_{(0)} \right)^2 } \right)  \Bigg)   \Bigg\}, \
\end{aligned}
\end{equation}
whose solution is
\begin{equation}\label{longsolutiongravity}
\begin{aligned}
{\OGG}_{ij}^{\text{inhom.}} &=\frac{4}{z}\left( \bar{g}_{(0)ik}\bar{g}_{(0)jl} - \frac{1}{d-1}\bar{g}_{(0)ij}\bar{g}_{(0)kl} \right)\, \times \\ \nonumber
&\quad \Bigg\{  \frac{ \left(b^2 g r - 2 \varphi \right)^2}{4gb^2 \left(\alpha-1 \right) } R_{\bar{g}_(0)}^{kl} + \frac{  b^2 g r - 2 \varphi}{gb^2\left( 2\alpha-1 \right) } D_{(0)}^{k}D_{(0)}^{l}\varphi_{(0)}\\ 
& \quad +\frac{ 1}{ g b^3 \alpha} D_{(0)}^{k} \varphi_{(0)} D_{(0)}^{l} \varphi_{(0)} + \frac{1-b^2}{ g b^4 \alpha} D_{(0)}^{k} \varphi_{(0)} D_{(0)}^{l} \varphi_{(0)} \\ \nonumber
& \quad -\frac{  \left(b^2 g r - 2 \varphi \right)^2}{8 g b^2 \left(\alpha - 1 \right) } \bar{g}_{(0)}^{kl}R[\bar{g}_{(0)}]  -\frac{1}{ g b^2 \alpha} \bar{g}_{(0)}^{kl} D_{(0)}^{m} \varphi_{(0)}D_{(0)m} \varphi_{(0)} \\
& \quad  - \frac{ b^2 g r - 2 \varphi }{2gb^2 \left(2\alpha-1 \right) } \Box_{\bar{g}_{(0)}} \varphi_{(0)}  + \frac{2 b^2-1 }{2 g b^4 \alpha } \bar{g}_{(0)}^{kl} D_{(0)}^{m} \varphi_{(0)}D_{(0)m} \varphi_{(0)}  \Bigg\} \
\end{aligned}
\end{equation}
It is not very hard to check that the above equation is indeed subleading to the 0-order solution we found earlier. Therefore, up to $n=1$, the asymptotic expansion of the bulk-induced metric near the boundary takes the form
\begin{align}
    \gamma{ij} = {\OG}_{ij}^{\text{inhom.}} + {\OGG}_{ij}^{\text{inhom.}} + \text{higher order terms}.
\end{align}

We can proceed with the expansion of the dilaton at order $n=1$. The same process must be followed for the dilaton flow equation. We have
\begin{equation}
\dot{\varphi} = \frac{\kappa^2}{\sqrt{-\gamma}} \frac{\delta}{\delta \varphi}(\CS_{(0)} + \CS_{(2)} + \ldots),
\end{equation}
and using the definition of the canonical momenta
\begin{equation}
\dot{\varphi} = \frac{\kappa^2}{\sqrt{-\gamma}} (\pi_{(0)\varphi} + \pi_{(2)\varphi} + \ldots),
\end{equation}
or, by expanding out $\pi_{(0)\varphi}$ we get
\begin{equation}
\dot{\overset{(0)}{\varphi}} + \dot{\overset{(2)}{\varphi}} + \ldots = \frac{\kappa^2}{\sqrt{-\gamma}} (\pi_{(0)\varphi} + \frac{b}{2}\pi_{(0)\varphi} \overset{(2)}{\varphi} + \ldots + \pi_{(2)\varphi} + \ldots),
\end{equation}
where again we cannot just naively use $\pi_{(2)\varphi}$. Thus, for $n=1$, we have to solve
\begin{equation}
  \dot{\overset{(2)}{\varphi}}  = \frac{\kappa^2}{\sqrt{-\gamma}} \left( \frac{b}{2}\pi_{\varphi(0)} \overset{(2)}{\varphi} + \pi_{\varphi(2)} \right),
\end{equation}
where the two canonical momenta shown above are given by
\begin{equation}
\begin{aligned}
\pi_{(0)\varphi} &= \frac{\sqrt{-\gamma}}{\kappa^2} g\frac{b}{2} e^{\frac{b}{2}\overset{(0)}{\varphi}} =\frac{\sqrt{-\gamma}}{\kappa^2}  \frac{1}{4}gb^2 \left( -\frac{1}{2}gb^2 r + \varphi{(0)} \right)^{-1} , \\
\pi_{(2)\varphi}&=\frac{\sqrt{-\gamma}}{\kappa^2}(\Xi' R + M' \partial_i \varphi \partial^i \varphi - M\Box \varphi) \
\end{aligned}
\end{equation}
The task is to solve the inhomogeneous equation
\begin{equation*}
\frac{d \overset{(2)} \varphi }{dr} =  \frac{1}{4}gb^2 \left( -\frac{1}{4}gb^2 r + \varphi_{(0)} \right)^{-1}\overset{(2)} \varphi + (\Xi' R + M' \partial_i \varphi \partial_i^i \varphi - M\Box \varphi).
\end{equation*}
The inhomogeneous solution is given by
\begin{equation}
{\overset{(2)}{\varphi}} = e^{-I(r)} \int e^{I(r)} (\Xi' R + M' \partial_i \varphi \partial_i^i \varphi - M\Box \varphi),
\end{equation}
where 
\begin{equation}
I(r) = -\log\left( -\frac{1}{2} g b^2 r + \varphi_{0} \right).
\end{equation}
Thus, we have
\begin{equation}
\begin{aligned}
{\overset{(2)}{\varphi}} &= \frac{1}{z}e^{-I(r)} \int dr\, e^{I(r)} e^{-\frac{b}{2}} e^{-2\bar{A}} \Bigg\{ -\frac{b}{2} R + \frac{b}{4} \frac{4 D_{(0)m}  \varphi_{(0)} D_{(0)}^{m}  \varphi_{(0)}}{b^2(-\frac{1}{2} g b^2 r + \varphi_{(0)}) } \\
                        & \qquad + \frac{1}{2} \left( \frac{2  D_{(0)m}  \varphi_{(0)} D_{(0)}^{m}  \varphi_{(0)} }{b(-\frac{1}{2} g b^2 r + \varphi_{(0)})^2} \right) -          \frac{2D_{(0)m} D_{(0)}^{m}  \varphi_{(0)} }{ b(-\frac{1}{2} g b^2 r + \varphi_{(0)}) }  \Bigg\}.
\end{aligned}
\end{equation}
Note that the term $e^{-2\bar{A}}$ comes from the fact that each of the tensors contains an inverse metric. Thus, this term is required in order to write our expression in terms of $\bar{g}_{(0)ij}$. Also, we find the following 
\begin{equation}
e^I(r) = \left( -\frac{1}{4} g b^2 r + \varphi_{0} \right)^{-1},
\end{equation}
\begin{equation}
e^{-I(r)} = \left( -\frac{1}{4} g b^2 r + \varphi_{0} \right),
\end{equation}
and
\begin{equation}
e^{-\frac{b}{2}\varphi} = e^{-\frac{b}{2} \left(-\frac{2}{b} \log(-\frac{1}{4}gb^2 r + \varphi_{(0)}) \right)  } = \left( -\frac{1}{4} g b^2 r + \varphi_{0} \right) = e^{-I(r)}
\end{equation}
Using the above equation we have
\begin{equation}
\begin{aligned}
{\overset{(2)}{\varphi}} &= \frac{1}{z} \left( -\frac{1}{2} g b^2 r + \varphi_{0} \right)  \int dr\, e^{-2\bar{A}} \Bigg\{ -\frac{b}{2} R_{\bar{g}_{(0)}} + \frac{b}{4} \frac{4 D_{(0)m}  \varphi_{(0)} D_{(0)}^{m}  \varphi_{(0)}}{b^2(-\frac{1}{2} g b^2 r + \varphi_{(0)}) } \\
& \qquad + \frac{1}{2} \Bigg( \frac{2  D_{(0)m}  \varphi_{(0)} D_{(0)}^{m}  \varphi_{(0)} }{b(-\frac{1}{2} g b^2 r + \varphi_{(0)})^2} - \frac{2D_{(0)m} D_{(0)}^{m}  \varphi_{(0)} }{ b(-\frac{1}{4} g b^2 r + \varphi_{(0)})  }\Bigg)   \Bigg\}.
\end{aligned}
\end{equation}
We can substitute the explicit form of $\bar{A}(r)$ in our expression. Recall that
\begin{equation}
\bar{A}(r) = \frac{2}{b^2(d-1)} \log \left(-\frac{1}{2}gb^2r + \varphi_{(0)} \right) = \frac{\alpha}{2}\log \left(-\frac{1}{4}gb^2r + \varphi_{(0)} \right) 
\end{equation}
according to our definition of $\alpha = \frac{4}{b^2(d-1)}$ we saw previously. Therefore, the exponential inside the integral becomes
\begin{equation}
e^{-2\bar{A}(r)} =\left(-\frac{1}{2}gb^2r + \varphi_{(0)} \right)^{-\alpha} 
\end{equation}
Thus, we have to perform the following integration
\begin{equation}
\begin{aligned}
{\overset{(2)}{\varphi}} &= \frac{1}{z} \left( -\frac{1}{2} g b^2 r + \varphi_{0} \right)  \int dr\, \frac{1}{\left(-\frac{1}{2}gb^2r + \varphi_{(0)} \right)^{{\alpha}} } \Bigg\{ -\frac{b}{2} R_{\bar{g}_{(0)}} \\
&\quad +  \frac{2 D_{(0)m}  \varphi_{(0)} D_{(0)}^{m}  \varphi_{(0)}}{b(-\frac{1}{2} g b^2 r + \varphi_{(0)})^2 }  - \frac{D_{(0)m} D_{(0)}^{m}  \varphi_{(0)} }{ b(-\frac{1}{2} g b^2 r + \varphi_{(0)})  }  \Bigg\},
\end{aligned}
\end{equation}
whose solution is
\begin{equation}
\begin{aligned}
{\overset{(2)}{\varphi}} &=-\frac{\left(-\frac{1}{2} b^2 g r+\varphi \right)^{2-a}}{(a-1) b g z} R_{\bar{g}_{(0)}} + \frac{4 \left(-\frac{1}{2} b^2 g r+\varphi \right)^{-a}}{(a+1) b^3 g z} D_{(0)m} \varphi_{(0)} D_{(0)}^m \varphi_{(0)} \\
                      &\qquad-\frac{2 \left(-\frac{1}{2} b^2 g r+\varphi \right)^{1-a}}{a b^3 g z} D_{(0)m} D_{(0)}^m \varphi_{(0)}. 
\end{aligned}
\end{equation}
Therefore, up to order $n=1$ the asymptotic form of the dilaton near the boundary takes the form
\begin{align}
    \varphi = \overset{(0)}{\varphi}(r,x) + \overset{(2)}{\varphi}(r,x) + \text{higher order terms},
\end{align}

The contribution of the gauge field, as we mentioned earlier, is trivial but irrelevant for our purpose. We only get a contribution from the gauge field at $n=2$, thus for $\pi_{(4)}^i$. Therefore, we simply have
\begin{equation}
{\overset{(2)}{A_i}} = {\overset{(2)}{A_i}}(x).
\end{equation}

Before we continue, we must make sure that both 2nd-order inhomogeneous solutions that we derived are subleading in the asymptotic expansion. We see that in Eq. \eqref{1416} the radial coordinate $r$ is raised to the power $\alpha$. Comparing with Eq. \eqref{longsolutiongravity} we see that in order the latter to be subleading, it is required that $\alpha > 2$. The same holds true for the equivalent dilaton contributions. Recall that we defined this constant $\alpha$ as 
\begin{equation*}
\alpha = \frac{4}{b^2(d-1)}.
\end{equation*} 
We can only draw some conclusions and get some understanding by knowing what $b$ is in the above equation. In Ref. \cite{Kanitscheider_2008}the authors set up a system, much similar to ours but without a gauge field in the action to begin with, and performed holography. They began with the dual (string-like) frame action for the non-conformal branes in ten dimensions
\begin{align}
    S = -\frac{N^2}{(2\pi)^7\alpha'^4} \int d^{10}x \, \sqrt{g}  e^{\gamma \varphi}\left(R + \beta(\partial \varphi)^2 - \frac{1}{2(8-p)!N^2}(F_{8-p})^2  \right),  
\end{align}
where, for $p<3$, $F_{p+2} = \star F_{8-p}$ is the (Hodge) dual field strength. For $p \neq 5$ the field equations in the string frame admit an $\AdS_{p+2} \times \mathbb{S}^{8-p}$ solution with linear dilaton, as we discussed in the introduction. By reducing the field equations over the sphere, it is possible to truncate to the $(p+2)=(d+1)$-dimensional metric $\tilde{g}_{\mu \nu}$ and dilaton $\tilde{\varphi}$. The ansatz for this truncation is
\begin{equation}
\begin{aligned}
ds_{\text{string}}^2 &= \alpha' d_p^{-c}(\CR^2 \tilde{g}_{\mu \nu} dx^{\mu} dx^{\nu} + d\Omega_{8-p}^{2}), \\
e^{\varphi} &= g_s(r_0^2\CR^2)^{ \frac{ (p-3)(7-p) }{4(5-p)   }e^{\tilde{\varphi}}}, \
\end{aligned}
\end{equation}
where $r_0^{7-p} \equiv Q_p$ and $\CR$ is defined below. From this we see that the equations of motion of the lower-dimensional fields for the $Dp$-branes follow from an action of the form
\begin{align}
   S = -L \int d^{D}x \, \sqrt{g} \, e^{\gamma \varphi} (R + \beta(\partial \varphi)^2 + C), 
\end{align}
where the constants $L, \beta, \gamma,$ and $C$ all depend on the specific system we are interested in. That is, they depend on $p=d+1$ of the $Dp$-branes responsible for the bulk theory. The constants of interest, for us, are
\begin{eqnarray*}
 \gamma = \frac{2(p-3)}{7-p}, && \beta = \frac{ 4(p-1)(p-4) }{ (7-p)^2 }, \\
\CR = \frac{2}{5-p}, && C = \frac{1}{2}(9-p)(7-p)\CR^2. \
\end{eqnarray*}
Notice that this action is in the so-called string frame. It is multiplied by an overall scale factor $e^{\gamma \varphi}$. In order to compare it with our action, we have to take it to the Einstein frame by a Weyl transformation. In the Weyl frame, the previous action can be written as
\begin{align}
    \tilde{S} = -L \int d^Dx \, \sqrt{g} \left( R - \frac{1}{2}(\partial \tilde{\varphi})^2 + Ce^{ \sqrt{ \frac{2(3-p)^2}{p(9-p)} }\tilde{\varphi} } \right),
\end{align}
where $\tilde{\varphi} = \left(\frac{8(9-p)}{p(7-p)^2}\right)^{1/2} \varphi $ and we find that
\begin{align}
    b^2 = \frac{2(3-p)^2}{p(9-p)}.
\end{align}
Therefore, the coefficient $\alpha$ that appears in the powers of various terms in the asymptotic expansions takes the form, as a function of $p$,
\begin{align}
\alpha(p) = \frac{2(9-p)}{(3-p)^2}.
\end{align}
Now important conclusions can be drawn. First, recall that we require $\alpha >2$. This constraint is satisfied for $p=0,1,2,4,5$. For $p=3$ the coefficient $\alpha$ diverges, while for $p=6,7,8,9$ the constraint is not satisfied. This means that our recursive method definitely cannot work for the cases where the constraint is not satisfied. We have successfully used the recursive method for $p=1,2$ however it can also be done for $p=0,4$ as well. The systems with $D3$-branes, though, asymptotically give an $\AdS$ geometry, as opposed to the conformally asymptotically $\AdS$ geometry of the non-conformal branes, and their dual theory is a conformal theory, namely its $\beta$-function vanishes. The case of $D5$-branes, which apparently satisfies our constraint, is quite different. The five-branes behave in a different way compared to, say $D2$-branes or $D3$-branes since they give a background geometry $E^{5,1}\times \mathbb{R} \times \mathbb{S}^3$ which is a non-$\AdS$ background. They involve linear dilatons, and the qualitative picture of holography for both $D5$-branes and $NS5$-branes is different. For such kind of systems see \cite{Karch:2005ms} and references therein.

\subsection{One-point functions and Ward identities}
Let us begin this subsection by recalling some standard notions regarding correlation functions and Ward identities of QFTs.

Ward identities are written in terms of the generating functional of correlation functions $W_{\text{QFT}}$. Recall that for a scalar field theory,
\begin{align}
    Z =  \int [D\varphi] e^{-S[\varphi] - \int d^dx \, \varphi(x) \CO(x)},
\end{align}
where $\CO(\varphi)$ is the composite operator of scalar fields whose source is $J(x) = \varphi(x)$, we can compute connected correlation functions considering derivatives of
\begin{align}
W_{\text{QFT}}  = \log Z 
\end{align}
with respect to the source $\varphi$ and then taking the limit where $\varphi \to 0$. One-point functions of the scalar field composite operator can be written as
\begin{align}
    \braket{ \CO(x) } = -\frac{\delta W_{\text{QFT}}}{\delta \varphi(x)} \Bigg|_{\varphi(x)\to 0} =  -\frac{\delta W_{\text{QFT}}}{\delta \varphi_{(0)}(x)} 
\end{align}
Similarly, for more general gauge theories, on arbitrary backgrounds, we have 
\begin{eqnarray}
\braket{ T_{ij}(x)} &=&   - \frac{2}{\sqrt{g_{(0)}}}  \frac{\delta W_{\text{QFT}}}{\delta g_{(0)}^{ij}(x)} \\
\braket{J_{i}(x) } &=&   - \frac{1}{\sqrt{g_{(0)}}}  \frac{\delta W_{\text{QFT}}}{\delta A_{(0)}^{i}(x)} \\
\braket{ \CO(x) } &=& -\frac{1}{\sqrt{g_{(0)}}}  \frac{\delta W_{\text{QFT}}}{\delta \varphi_{(0)}(x)}
\end{eqnarray}
To compute higher-order correlation functions, we need to further differentiate with respect to the sources and then set them to zero. The divergences that might appear are removed by renormalization. We can use the bulk boundary term, which is identified with the renormalized generating functional of the correlation functions in accordance with the holographic dictionary, to define the counter-terms and a renormalization scheme.

Further recall that the Ward identities can be expressed in terms of the one-point functions in the presence of sources. For example, to the classically conserved current $J^i$, corresponds to the fact that under infinitesimal gauge transformations $W_{\text{QFT}}$ remains invariant, that is, $\delta{A} W_{\text{QFT}}=0$ leading to the Ward identity
\begin{align}
    D_i \braket{ J_{i}(x) } = 0.
\end{align}
If we consider the invariance under coordinate transformations, the corresponding Ward identity is
\begin{align}
    D^i\braket{ T_{ij}(x) } - \braket{J_{i}(x) }  F_{(0)ij}(x) + \braket{ \CO(x)D_j \varphi_{(0)}(x) } = 0.
\end{align}
We can consider what happens in the case of conformal transformation under which the sources transform as
\begin{eqnarray*}
\delta g_{(0)ij} &=& 2  \delta {\sigma}(x) g_{ij}, \\
\delta A_{(0)i} &=& 0, \\
\delta \varphi_{(0)} &=& -(d-\Delta) \delta {\sigma}(x) \
\end{eqnarray*}
The generating functional of correlation functions does not necessarily have to be invariant under conformal transformations. Actually, by varying the generating functional with respect to conformal transformations, we obtain
\begin{align}
    \delta W = \int d^dx \, \sqrt{g_{(0)}} \delta \sigma(x) \CA,
\end{align}
that is, the conformal anomaly, where $\CA$ is the anomaly density, a local functional of the sources. The conformal rescaling gives the trace Ward identity
\begin{align}
    \braket{ T_i^{\,\, i}(x) } = -(d-\Delta)\phi_{(0)} \braket{ \CO(x)} + \CA.
\end{align}
$W_{\text{QFT}}$ is not invariant nor under scale transformation and the reason behind this is that renormalization introduces a scale parameter on which the generating functional depends upon. What is invariant under scale transformations, on the other hand, is the following equation
\begin{align}
    \mu \frac{\partial}{\partial \mu} W_{\text{QFT}} = \int d^dx \left( 2g_{(0)ij} \frac{\delta}{\delta g_{(0)ij}} + (\Delta - d)\phi_{(0)} \frac{\delta}{\delta \phi_{(0)}} \right),
\end{align}
where $\mu$ is the scale parameter and is related to the constant $\sigma$ as $\mu = e^{\sigma}$.

In the context of holography, the one-point functions are identified as the renormalized momenta of the theory, which are given by the modes for which the iterative algorithm breaks. When we hit the first finite term in the expansion $\CS = \CS_{(0)} + \CS_{(2)} + \ldots + \CS_{(d)} + \ldots$, there exists a new solution, which is independent of the Hamilton-Jacobi equation, in that order, providing the one-point functions. We saw previously how the fields are expanded asymptotically. Also, recall that earlier we made the identification
\begin{equation*}
S_b \Big|_{r_0} = \CS \big|_{r_0},
\end{equation*}
and this boundary term is identified to be the term containing all the divergences. Then the renormalized action is given by the sum of the regulated action and the counter-term action
\begin{equation}
S_{\text{ren.}} = \lim_{r\to \infty} (S + S_{\text{ct.}} ),
\end{equation}
and, as we discussed earlier, the $\AdS/\CFT$ dictionary identifies the renormalized action with the generating functional of the renormalized correlation functions of the dual theory. Thus, the one-point functions will be given by one functional differentiation with respect to the corresponding source. This differentiation amounts to the renormalized momenta $\pi_{(d)}$. Therefore, the one-point functions are given by
\begin{eqnarray}
\braket{ \mathcal{T}^{ij} }_{\text{ren.}} &=& \frac{\delta S_{\text{ren.}}}{\delta \gamma_{ij}} \,\,\,=\,\,\, \pi_{(d)}^{ij} \\
\braket{ \mathcal{J}^{i} }_{\text{ren.}}&=& \frac{\delta S_{\text{ren.}}}{\delta A_{i}} \,\,\,=\,\,\, \pi_{(d)}^{i} \\
\braket{ \CO }_{\text{ren.}} &=& \frac{\delta S_{\text{ren.}}}{\delta \varphi} \,\,\,=\,\,\, \pi_{(d)}\
\end{eqnarray}
We recommend the reader to consult Ref. \cite{Papadimitriou2016} which provides a much deeper analysis on the above. 

Recall that $2n=d$ is the order of the term in which the recursive solution breaks down because then we get a zero eigenvalue of $\delta{\gamma}$. We do not discuss the explicit expressions for the case of $Dp$-branes for $p=1,2,4$, but this is the general way to obtain the one-point functions. Note that for theories of even dimensions, it has been shown \cite{deHaro:2000vlm} that there might exist a conformal anomaly as opposed to odd dimensions where no conformal anomaly appears, and the asymptotic expansion might include some logarithmic terms whose origin is traced to the fact that the trace of the stress-energy tensor of a conformal field theory on a curved background in even dimensions has a nonzero value and picks up this anomaly. For example, in even dimensions, the metric takes the form
\begin{align}
    g_{ij}(x,r) = g_{ij\,(0)}(x) + r^2 g_{ij\, (2)} + \ldots + r^d g_{ij\, (d)} + h_{ij \, (d)}r^d \log r^2 + \CO(r^{d+1}),
\end{align}
where the tensor $h_{ij \, (d)}$ of the logarithmic term is directly related to the conformal anomaly and in specific it is proportional to the metric variation of the conformal anomaly \cite{deHaro:2000vlm}. Therefore, the same kind of terms appear in the expansions of the conjugate momenta as well. Despite the fact that for $D1$-branes and $D2$-branes, where the iterative process breaks earlier than in the case of, say, $D4$-branes, no such terms appear, as can be seen from the asymptotic expansions of the induced fields.

Finally, as we discussed much earlier, in a generally covariant theory the constraints coming from the Hamilton-Jacobi formulation must identically vanish. Eq. \eqref{HamJac} was the Hamilton-Jacobi equation, which we solved and there were left another two constraints which are re-written here for convenience
\begin{eqnarray*}
\mathcal{H}_i    &=&   (-2D^{j})\pi_{ij}^{ } + \pi_{\varphi}\partial_i \varphi + \pi^j F_{ij}, \\ \nonumber
			&& \\ 
\mathcal{F}      &=& -D_i \pi^i. \
\end{eqnarray*}
In particular, by substituting the renormalized action into these constraints we obtain the holographic Ward identities,
which are well known \cite{Papadimitriou_2011}

\begin{eqnarray}
0 &=& -2D^j \braket{ \mathcal{T}_{ij} }_{\text{ren.}}  + \braket{\mathcal{O}}_{\text{ren.}} \partial_i \varphi + \braket{ \mathcal{J}^{j} }_{\text{ren.}} F_{ij}    , \\ 
0&=& D_i \braket{ \mathcal{J}^{i} }_{\text{ren.}}   .
\end{eqnarray}
Again, we do not proceed into the explicit calculation of for the specific cases for $p=1,2$ but the procedure is analytically sketched in a similar fashion in Ref. \cite{Papadimitriou2016}.

\section{Conclusions}
We have reviewed the material needed to understand the basics of holographic renormalization for non-conformal branes, using the radial Hamilton-Jacobi method, by providing explicit computations for the Einstein-Maxwell theory coupled to a dilaton. The theory we worked on is somewhat similar to the axion-dilaton theory of \cite{Papadimitriou_2011}, however, at the time of writing the author's Master thesis, no explicit computations for the former were available. We performed holographic renormalization using the radial Hamilton-Jacobi method and the resursive algorithm Papadimitriou \cite{Papadimitriou_2010} which provides the most generic framework for systematic holographic renormalization.

This is particularly interesting since non-conformal branes arise not only from the question of whether holography can be performed for non-exactly asymptotically $\AdS$ spaces, but also because the case of non-coformal branes are analogous to the one of the deformations of SYM theories where the asymptotic value of the dilaton determines the value of the dimensionful coupling constant of the gauge theory, which might provide further insight for non-conformal dual theories.

 We have reconfirmed that holography can be performed for non-conformal branes, and we have worked out how to obtain the holographic dictionary. Starting from the action functional of the theory in the bulk, performing the ADM decomposition, and focusing on radial slices near infinity, we have used an iterative algorithm that solves the radial Hamilton-Jacobi equation in an asymptotic manner giving us order-by-order solutions for backgrounds which asymptotically are $\AdS_{p+2}\times \mathbb{S}^{8-p}$. We have found that this method works for, $Dp$-branes with $p=0,1,2,4$ in accordance to \cite{Kanitscheider_2008}.  At the time of writing of the thesis, further checks that the one-point functions, which are given by the variation of the renormalized on-shell action with respect to the bulk fields and Ward identities, which are given by the constraints arising from the Hamiltonian formulation of the gravitational problem, are in full accordance between the two methods were required. Unfortunately, time constraints did not allow for this analysis, but our results, at least in a superficial level, look indeed consistent and verified by several papers published afterwards. Furthermore, when the method presented is applied directly to the dual frame, and not the sting-like frame, of the reduced-dimensional effective action, it seems that efficiency gains can be acquired since the covariant counter-term action is easier to derive and also because working with the canonical momenta instead of the on-shell action also seems to be a more efficient way to perform such computations. It is not difficult to proceed to explicit computations of $n$-point functions or other renormalized holographic observables.

However, the previous analysis might not be as straightforward as one might naively think when supersymmetric solutions are taken into consideration. It was found in \cite{Genolini2017} and also in the careful analysis of \cite{Papadimitriou:2017kzw} that holographic renormalization for supersymmetric solutions of certain supergravity theories. For example, for the the $d=5$, $\mathcal{N}=2$ minimal gauged supegravity, holographic renormalization is more ``tricky'' and subtle. We refer to those papers for further details. In a similar direction, we refer to Ref. \cite{An:2017ihs} for a systematic approach to supersymmetric holographic renormalization for generic $d=5$ $\mathcal{N}=2$ supergravity theories. These considerations have been important in the understanding of anomalies in the Ward identities of the dual SCFTs of these supergravity theories. We emphasize that holoraphic renormalization, based on \cite{Papadimitriou_2010}, reviewed in this paper, and extended for supersymmetric solutions, precisely on these $d=5$, $\mathcal{N}=2$ supergravity theories will aid to a further understanding on their very interesting $d=4$, $\mathcal{N}=1$ dual SCFTs.

The radial Hamilton-Jacobi method is an important tool in the analysis of various related system and keeps being used in a variety of contexts, for example \cite{Rajagopal:2015lpa,Erdmenger:2016jjg,Elvang:2016tzz,Chen:2019zlg,Kim:2020dqx,Anastasiou:2020zwc,Chandrasekaran:2021vyu,Aniceto:2021xhb,Santos:2022zvu} to name a few.

\acknowledgments

The author wishes to acknowledge the generous support of the OP RDE funded project CZ.02.1.01/0.0/0.0/16 019/0000765 ``Research Center for Informatics''. 
The author also acknowledges the support of the 2012-2013 and 2013-2014 ``UAM Campus de Excelencia'' awards.


\bibliography{refs}
\bibliographystyle{JHEP}  
\end{document}